\documentclass[10pt]{article}
\usepackage[margin=1in]{geometry}
\usepackage[utf8]{inputenc}
\usepackage{graphicx}
\usepackage{xcolor}
\usepackage{soul}
\usepackage{caption}
\usepackage{subcaption}
\usepackage{multirow}
\usepackage{indentfirst}
\usepackage{booktabs}
\usepackage{multirow}
\usepackage[colorlinks=true,linkcolor=blue,citecolor=blue]{hyperref}
\usepackage{bm}
\usepackage{float}
\usepackage[sort,square,numbers]{natbib}
\usepackage[noblocks]{authblk}
\usepackage{amsmath, amsthm, amssymb, amsfonts}
\definecolor{jlcol}{rgb}{255, 0, 0}
\definecolor{descol}{rgb}{0, 0, 255}

\title{A Probabilistic Neural Twin for Treatment Planning in Peripheral Pulmonary Artery Stenosis}

\author[1]{John D. Lee}
\author[2]{Jakob Richter}
\author[2]{Martin~R.~Pfaller}
\author[2]{Jason M. Szafron}
\author[2]{Karthik Menon}
\author[2]{Andrea Zanoni}
\author[6]{Michael R. Ma}
\author[2,3]{Jeffrey A. Feinstein}
\author[5]{Jacqueline Kreutzer}
\author[2,3,4]{Alison~L.~Marsden}
\author[1,*]{Daniele~E.~Schiavazzi}

\affil[1]{Department of Applied and Computational Mathematics and Statistics, University of Notre Dame, Notre Dame, IN 46556, US.}
\affil[2]{Department of Pediatrics (Cardiology), Stanford University, CA 94305, US.}
\affil[3]{Department of Bioengineering, Stanford University, CA 94305, US.}
\affil[4]{Institute for Computational and Mathematical Engineering, Stanford University, CA 94305, US.}
\affil[5]{Department of Pediatrics, University of Pittsburgh School of Medicine and UPMC Children’s Hospital of Pittsburgh, PA 15213, US.}
\affil[6]{Department of Cardiothoracic Surgery, Stanford University,
870 Quarry Rd, Stanford, CA 94304, US.}
\affil[*]{Corresponding Author. Email: dschiavazzi@nd.edu}

\date{\today}

\begin{document} 

\maketitle

\begin{abstract}
\noindent The substantial computational cost of high-fidelity models in numerical hemodynamics has, so far, relegated their use mainly to \emph{offline} treatment planning.
New breakthroughs in data-driven architectures and optimization techniques for fast surrogate modeling provide an exciting opportunity to overcome these limitations, enabling the use of such technology for time-critical decisions. 
We discuss an application to the repair of multiple stenosis in peripheral pulmonary artery disease through either transcatheter pulmonary artery rehabilitation or surgery, where it is of interest to achieve desired pressures and flows at specific locations in the pulmonary artery tree, while minimizing the risk for the patient.
Since different degrees of success can be achieved in practice during treatment, we formulate the problem in probability, and solve it through a sample-based approach.
We propose a new offline-online pipeline for probabilsitic real-time treatment planning which combines \emph{offline} assimilation of boundary conditions, model reduction, and training dataset generation with \emph{online} estimation of marginal probabilities, possibly conditioned on the degree of augmentation observed in already repaired lesions.
Moreover, we propose a new approach for the parametrization of arbitrarily shaped vascular repairs through iterative corrections of a zero-dimensional approximant.
We demonstrate this pipeline for a diseased model of the pulmonary artery tree available through the Vascular Model Repository.
\end{abstract}

\section{Introduction}\label{sec:intro}
%
\noindent Peripheral pulmonary artery stenosis (PPAS) is a condition occurring most frequently in conjunction with congenital heart diseases such as Williams or Allagile Syndrome~\cite{virtual_intervention}, or less commonly after surgical interventions or as an isolated congenital condition~\cite{Inglessis_Landzberg_2007}. The varying comorbidities and complex positioning of stenosis continue to pose significant challenges in the diagnosis and treatment of PPAS~\cite{TRIVEDI_BENSON_2003}.
Previous clinical studies have shown that surgical pulmonary artery reconstruction or patch augmentation, a technique utilizing experience to ``sculpt'' a new pulmonary architecture, achieves long-term reduction of right ventricular pressures with low rates of morbidity, mortality, and reintervention~\cite{Collins_2019, Luong_2020}. 
However, due to significant technical challenges associated with the surgery, transcatheter interventions remain standard-of-care for pulmonary artery stenosis at most institutions despite sub-optimal outcomes that not only yield persistent right ventricular hypertension but also frequently necessitate re-intervention~\cite{Hallbergson_2013, Cunningham_2013}. 
In cath-based treatments of pulmonary artery stenosis with multiple lesions, clinicians must prioritize which lesions to balloon and/or stent and in which order. 
Even advanced procedures such as patch augmentation require the prioritization of lesion repairs, efficiently fixing the most impactful stenosis in a limited time, thus minimizing the risk for the patient. 
A pressing need thus exists to develop new predictive virtual treatment planning tools to aid clinicians in identifying optimal repair strategies.

Digital twins, virtual models representing physical processes, are widely used in many engineering tasks and have been shown as powerful resources in healthcare. 
For example, Bj\"{o}rnsson \textit{et. al.} designed personalized digital twins to exhaustively search the optimal drug for an individual~\cite{Bjornsson_2019}.
The first digital twins in the surgical guidance space date as early as 1994, when Taylor \textit{et. al.} developed CT-based pre-operative planning and a robotic guidance system for total hip replacement surgery, resulting in increased implant fit and placement accuracy~\cite{Taylor_1994}.
In recent years, significant research has been dedicated to developing digital twins for both pre- and peri-operative surgical usage in a variety of contexts~\cite{Coelho_2020, Shu_2023}.

In the context of PPAS, Lan \textit{et. al.} investigated peripheral pulmonary artery stenosis in six Williams and Alagille patients, constructing three-dimensional computational fluid dynamic (CFD) models to perform virtual transcatheter interventions~\cite{virtual_intervention}. 
Validating patient-specific hemodynamics based on available clinical measurements, Lan \textit{et. al.} demonstrated the potential for CFD models to serve as digital twins. However, several limitations acknowledged in Lan \textit{et. al.}'s study inhibited three-dimensional CFD models from being integrated in protocols for time-critical treatment decisions. 
First, virtually constructed interventions assumed guaranteed success of the procedure, whereas, in reality, each repair has an associated probability of success~\cite{Geggel_Gauvreau_Lock_2001}. 
The probabilistic nature of interventions, as well as the desire for real-time hemodynamic information, both imply a need for efficient CFD models. Nonetheless, the three-dimensional CFD models used in Lan \textit{et. al.}'s study require hours of computation on high-performance hardware, posing another significant challenge.

In this study, we address these limitations, presenting a prototype for a probabilistic treatment planning tool combining an offline cardiovascular modeling pipeline with the online real-time interrogation of data-driven surrogates, linking the degree of repair at multiple stenoses with the resulting pressures and flows at a number of pre-specified locations.
This appears to be a new direction for the use of cardiovascular digital twins during transcatheter intervention or surgery.

The paper is organized as follows. 
We begin, in Section~\ref{sec:prob}, by discussing a probabilistic framework combining probabilistic stenosis repair and numerical simulations to formulate a joint probability distribution fully characterizing post-treatment hemodynamics.
A patient-specific pulmonary anatomy and corresponding three-dimensional hemodynamic model are presented in Section~\ref{sec:threed_modeling}, while Section~\ref{sec:rom} introduces two zero-dimensional lumped parameter approximations, the first used for boundary condition tuning (Section~\ref{sec:bc_tuning}) and the second used as a three-dimensional model surrogate (Section~\ref{sec:lpn}).
A general framework is then introduced to iteratively correct the resistance distribution in zero-dimensional reduced order models to match any desired distribution of flows and pressures along a collection of vessel centerlines. This approach is used to generate zero-dimensional approximations of pre-treatment hemodynamics in Section~\ref{sec:linear_correction} and hemodynamics following intervention for three stenotic lesions in Section~\ref{sec:parameterization}.
A neural twin is then constructed by first generating datasets (Section~\ref{sec:training_gen}) to train a fully connected neural architecture (Section~\ref{sec:ann}), with resulting accuracies discussed in Section~\ref{sec:model_acc}.
By integrating the neural twin with the proposed probabilistic framework, we formulate the joint, and any conditional or marginal probabilities associated with the outcome of treatment. In Section~\ref{sec:results}, specific results from our sample model are used to illustrate how relevant clinical questions might be answered in a pre-operative and peri-operative scenario.
We conclude the paper with a discussion in Section~\ref{sec:discussion}.

\section{Probabilistic analysis of post-repair hemodynamics}\label{sec:prob}

\noindent In this section, we aim to provide a probabilistic perspective on virtual treatment for peripheral pulmonary artery disease. Our main goal is to quantify the sampling framework for the joint probability distribution $\rho(\bm{p},\bm{f},\bm{c})$, which represents the pressures $\bm{p}$, flow $\bm{f}$, and amount of repair $\bm{c}$ at each stenotic lesion. By having complete knowledge of this distribution, we can formulate treatment decisions based on conditional or marginal probabilities derived from $\rho(\bm{p},\bm{f},\bm{c})$. This approach allows us to take into account the inherent uncertainties associated with repair and make more informed and optimal treatment decisions.

Assume $n$ candidate locations in the pulmonary artery tree are selected for treatment by a clinician, and at each location, a target post-repair vascular anatomy is also identified. Note that here, we use the concept of \emph{repaired/treated configuration} without any assumptions that stenosis is concentrated at a single cross-section. In other words, a \emph{repair} at location $i$ in the pulmonary artery tree is a possibly complex change in the three-dimensional geometry from the diseased to the repaired configuration, described by a single parameter $c_{i}$, where $c_{i}=0$ corresponds to the pre-treatment (or \emph{diseased}) anatomy and $c_{i}=1$ to the desired repair at $i$.
Therefore, the outcome of virtual treatment can be represented simply as a realization from the vector of identically distributed random variables $\bm{c} = (c_1,c_2,\dots,c_n)\in[0,1]^{n}$.
We also assume a simple hierarchical model~\cite{gelman2006data} for each $c_{i},\,\,i=1,\dots,n$ where variables are uniformly distributed over a range that is dependent on a single-trial multinomial repair success with states \emph{failed} (F), \emph{partial/moderate} (M) or \emph{successful repair} (S). Formally, we can write
\begin{equation}\label{equ:c_density}
c_i \sim \mathcal{U}(u_s, l_s),\,\text{with}\,\,s\sim m(\{k_F, k_M, k_S\};1,P_{F},P_{M},P_{S}),
\end{equation}
where $k_F+k_M+k_S = 1$, $P_{F}+P_{M}+P_{S}=1$ are the probabilities associated with each state, and $u_F=0$, $l_S=1$ to keep $c_i\in[0,1]$. Since $k_i,\,i=\{F,M,S\}$ are non negative integers, this is equivalent to $s$ being defined on three states with probability masses $P_{F}$, $P_{M}$, $P_{S}$, respectively.

In addition, when augmenting the cross-section area of a pulmonary branch either by placing a stent or performing balloon angioplasty, we assume each identified location for repair undergoes its own intervention. This leads to the clinically sound assumption that the outcome of a repair at a single location does not affect the probability of a successful repair occurring at any other location in the pulmonary artery vasculature.  
Altogether, the following decomposition for the joint probability of the repair variables can be assumed, i.e.
\begin{equation}\label{equ:indep_c}
\rho(c_1,\dots,c_n) = \prod_{i=1}^{n}\rho(c_i).
\end{equation}

\begin{figure}[ht!]
\centering
\includegraphics[width=0.3\textwidth]{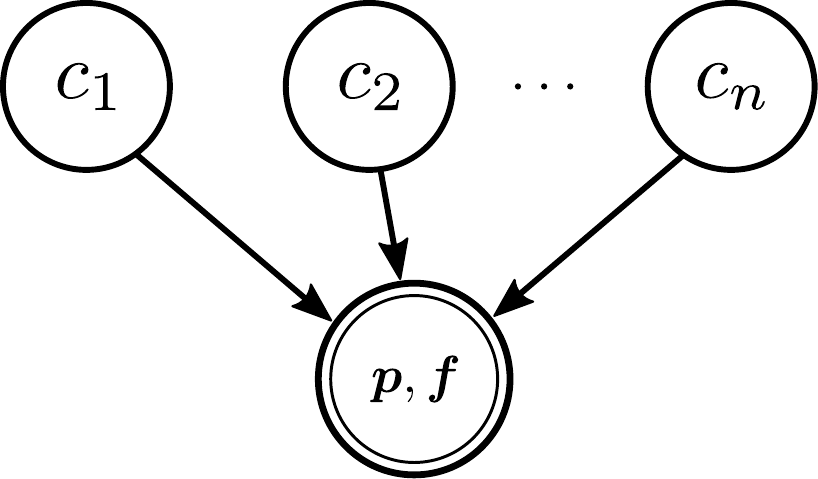}
\caption{Na\"ive Bayes network representation for joint density factorization in~\eqref{equ:joint_decompose}.}\label{fig:bn}
\end{figure}

We now turn our attention on how to assess the success of an intervention. Consider the pulmonary artery pressure and flow measured at $m_{p}$ and $m_{f}$ clinically relevant locations, quantified through the random vectors
\[
\bm{p} = (p_1,p_2,\dots,p_{m_{p}}), \bm{f} = (f_1,f_2,\dots,f_{m_{f}}).
\]
Note that these locations are different, in general, by the $n$ locations identified for treatment.
Pressures and flows at the $m_{p}$ and $m_{f}$ locations are subject to a zero-mean Gaussian measurement error with standard deviation $\sigma_{p}$ and $\sigma_{f}$, respectively.
As introduced above, our goal is to compute $\rho(\bm{p},\bm{f},\bm{c})$ and, from this joint probability distribution, compute all marginal and conditional probabilities to answer relevant clinical questions. 
Using the definition of conditional probability we can write
\begin{equation}\label{equ:joint_decompose}
\rho(\bm{p},\bm{f},\bm{c}) = \rho(\bm{p},\bm{f} \vert \bm{c})\rho(\bm{c}),
\end{equation}
and since the variables $c_i,\,\,i = 1,\dots,n$ are independent, we have
\begin{equation}\label{equ:joint}
\rho(\bm{p},\bm{f},\bm{c}) = \rho(\bm{p},\bm{f} \vert \bm{c})\cdot \prod_{i=1}^{n}\rho(c_{i}),\,\,\text{or}\,\,\log \rho(\bm{p},\bm{f},\bm{c}) = \log \rho(\bm{p},\bm{f} \vert \bm{c}) + \sum_{i=1}^{n}\log\rho(c_{i}).
\end{equation}
This decomposition of the joint probability distribution is represented through the simple na\"ive Bayes network in Figure~\ref{fig:bn}. 
Evaluation of the joint probability~\eqref{equ:joint} proceeds as a two-step process. The marginals $\rho(c_i),\,\,i=1,\dots,n$ are determined using the law of total probability
\[
\rho(c_{i}) = \sum_{j=\{F,M,S\}}\,\rho(c_{i}\vert u_{j},l_{j})\,\rho(u_{j},l_{j}) = \sum_{j=\{F,M,S\}}\,\dfrac{1}{l_{j}-u_{j}}\,P_{j}.
\]
The remaining factor is a conditional probability of the form
\begin{equation}
\rho(\bm{p},\bm{f}\vert \bm{c}) = \mathcal{N}\left(g(\bm{c}), \bm{\Sigma}_{p,f}\right),
\end{equation}
where a deterministic cardiovascular model $g:\mathbb{R}^{n}\rightarrow\mathbb{R}^{m_{p}+m_{f}}$ computes the pressures and flow rates at a number of locations of interest given a repaired configuration parameterized by $\bm{c}$, and the covariance 
\[
\bm{\Sigma}_{p,f} = 
\begin{bmatrix}
\bm{\sigma}_{p}^{2} &\\
& \bm{\sigma}_{f}^{2}\\
\end{bmatrix},
\]
where $\bm{\sigma}_{p}^{2}$ and $\bm{\sigma}_{f}^{2}$ are diagonal matrices of order $m_{p}$ and $m_{f}$, with diagonal entries containing the standard deviations for pressure and flow rate measurements, respectively.
In other words, for every realization of $\bm{c}$, the pressures $\bm{p}$ and flows $\bm{f}$ at a number of $m_{p}$ and $m_{f}$ locations can be computed through the solution of a zero-, one- or three-dimensional cardiovascular model $g$ and by adding measurement noise. 
We assume zero error for the three-dimensional model output, while the discrepancy of zero-dimensional models is minimized by iterative corrections, as discussed in Section~\ref{sec:linear_correction}.

In order to generate sufficiently accurate probability distributions corresponding to an $n$-dimensional sample space, tens of thousands of samples must be drawn and computed.
For the purposes of real-time guidance during transcatheter rehabilitation or surgery, even the lowest order zero-dimensional models are too computationally expensive, taking roughly 5 seconds on a desktop processor to solve each.
To significantly reduce such cost, we construct a fully connected neural network surrogate, trained on a set of precomputed solutions from $g(\bm{c})$. 
While network architectures for multifidelity model emulation and information fusion have recently been proposed in the literature (see, e.g.~\cite{meng2020composite}), in this paper we focus mainly on zero-dimensional models for generating the training data.
In the following sections, we discuss how, in practice, a repair geometry is parameterized and how the various components of the joint distribution are computed.

\section{Three-Dimensional Modeling}\label{sec:threed_modeling}

\subsection{Patient Anatomies}\label{sec:patient_anatomies}

%
\noindent We sourced a pulmonary vascular tree from the Vascular Model Repository (VMR, \url{https://www.vascularmodel.com}). This model originates from a female patient, aged 16.8 years, diagnosed with Allagile Syndrome (AS). To ensure privacy, all identifiable information from the model was removed.

The chosen diseased model corresponds to model AS-1 from~\cite{virtual_intervention} and was constructed starting from a computed tomography angiography (CTA) scan. 
We utilized SimVascular~\cite{updegrove2017simvascular} to trace path-lines and perform two-dimensional lumen segmentation on orthogonal cross-sections in both the main pulmonary artery and its branches, including up to five generations of branching.
The resulting segmentation was then lofted into a three-dimensional surface mesh which was used to generate a volume mesh. Additionally, virtual configurations following the post-treatment augmentation of the proximal left and right central pulmonary artery (LPA/RPA), and a distal region of the RPA were also generated under the supervision of an expert clinician, as discussed in~\cite{virtual_intervention}.
More detail on these \emph{repaired} configurations is available in Section~\ref{sec:parameterization}.

\begin{figure}[ht!]\newcommand\tw{0.065}
\centering
\begin{subfigure}[t]{0.4\linewidth}
    \includegraphics[width=\textwidth]{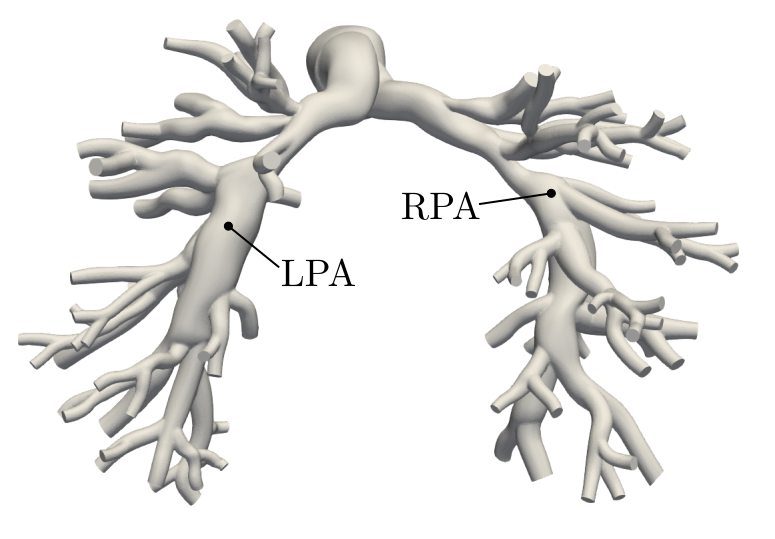}
    \caption{3D Model (VMR ID: AS1\_SU0308)}\
    \label{fig:AS1_SU0308_model}
\end{subfigure}
\begin{subfigure}[t]{0.4\linewidth}
    \includegraphics[width=\textwidth]{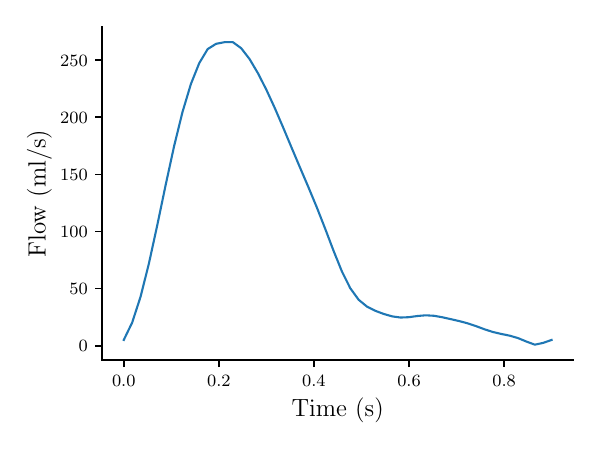}
    \caption{Inflow Waveform (VMR ID: AS1\_SU0308)}
    \label{fig:AS1_SU0308_inflow}
\end{subfigure}
\caption{Three-dimensional rendering and inlet flow waveform for the selected pulmonary artery anatomy.}
\label{fig:model_inflow}
\end{figure}

The diseased model was meshed with 6.67 million tetrahedral elements, and boundary conditions were applied at the inlet and 90 outlets. Three-dimensional model geometry and inflow are shown in Figure~\ref{fig:model_inflow}.
We were provided measurements for the main pulmonary artery (MPA), RPA, and LPA, detailing mean, systolic, and diastolic pressures. Other data included mean capillary wedge pressure, cardiac output, and flow splits between the left and right pulmonary trees. Exact values are reported in Table~\ref{table1}.
Although not directly available through the clinical records, we accessed an inflow waveform from the right heart lumped parameter model in~\cite{virtual_intervention} and adjusted it for consistency with the measured cardiac output. 
More information on this can be found in Appendix~\ref{sec:app_nonlin_res}.

\subsection{Hemodynamic Simulations}\label{cfd}

\noindent Rigid wall hemodynamic simulations were carried out using svSolver, a streamlined upwind Petrov–Galerkin (SUPG) stabilized incompressible Navier-Stokes finite element solver available through the SimVascular software platform~\cite{whiting2001stabilized}. 
The open-loop boundary conditions included a prescribed flow waveform (see Figure~\ref{fig:AS1_SU0308_inflow}) at the main pulmonary artery inlet, and three-element RCR Windkessel boundary conditions at each outlet. Additionally, the pulmonary capillary wedge pressure was prescribed as a constant left atrial pressure at all outlets. 
All simulations were run for 6 cardiac cycles to ensure a periodic hemodynamic response. 
A previous study~\cite{virtual_intervention} considered one-way fluid structure interaction through the Coupled Momentum method~\cite{figueroa2006coupled}, with vascular walls treated as an elastic membrane with out-of-plane shear stiffness, and only translational degrees of freedom. A comparison between the hemodynamic response of rigid and deformable wall models for the pulmonary arterial tree is discussed, for example, in~\cite{liu2020fluid} using models with open loop boundary conditions (prescribed MPA flow and three-element Windkessel outlet circuits). This study shows the tendency of rigid models to overestimate MPA pressures. However, we use MPA pressures as targets for boundary condition tuning, further reducing these differences. Future studies will further investigate how inclusion of fluid-structure interaction affects the prioritization of pulmonary stenosis repair.
\begin{table}
\centering
\caption{Available and relevant biometric, pressure and flow measurements for diseased (pre-operative) model.}
\begin{tabular}{l c l c}
\toprule
{\bf Measurement} & {\bf Value} & {\bf Cath. pressure} & {\bf Value}\\
\midrule
{\bf Age [yrs]} & 16.8 & {\bf MPA mean pressure [mmHg]}  & 42\\
{\bf Sex}  & Female & {\bf MPA systolic pressure [mmHg]}  & 90\\
{\bf CO [L/min]}  & 5.8 & {\bf MPA diastolic pressure [mmHg]}  & 18\\
{\bf PA Flow Split (R/L)}  & 52\%/48\% & {\bf Wedge pressure [mmHg]}  & 20 \\
\bottomrule
\end{tabular}
\label{table1}
\end{table}

\section{Reduced-Order Modeling}\label{sec:rom}

\noindent Due to the intractable computational costs involved in working with three-dimensional cardiovascular models, the use of reduced-order models is absolutely necessary to achieve the large amounts of hemodynamic results necessary for our pipeline.
In this section, we outline a comprehensive approach for creating two different zero-dimensional (i.e., circuit) reduced-order models: a simple surrogate to facilitate boundary condition tuning and a lumped parameter network (LPN) designed to approximate the hemodynamic response of the three-dimensional model. 
The following sections describe these two models in greater detail. 

\subsection{Boundary condition tuning}\label{sec:bc_tuning}

\begin{figure}[ht!]
\centering
\includegraphics[width=0.5\linewidth]{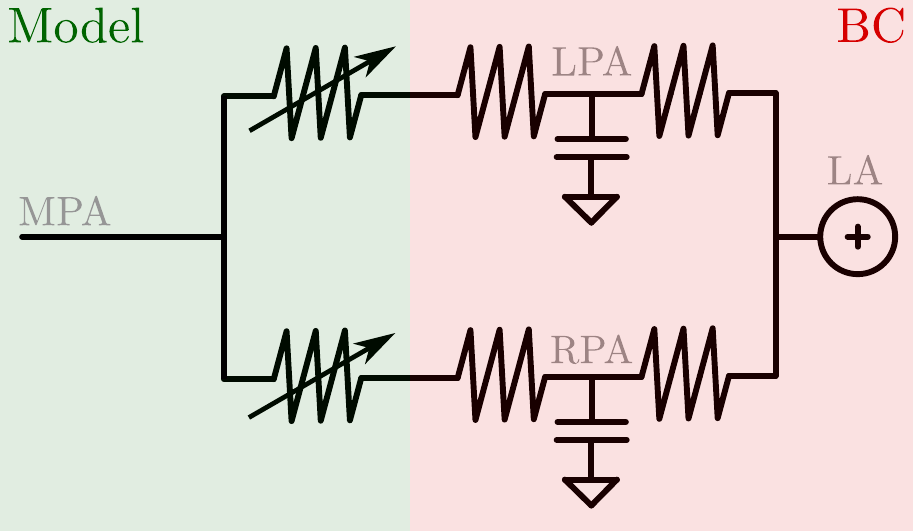}
\caption{Simplified circuit model for tuning RCR boundary conditions. Green indicates fixed values during optimization, whereas red contains the parameters to be optimized.}\label{fig:tuning_model}
\end{figure}

\begin{figure}[ht!]\newcommand\tw{0.065}
    \includegraphics[width=1.0\textwidth]{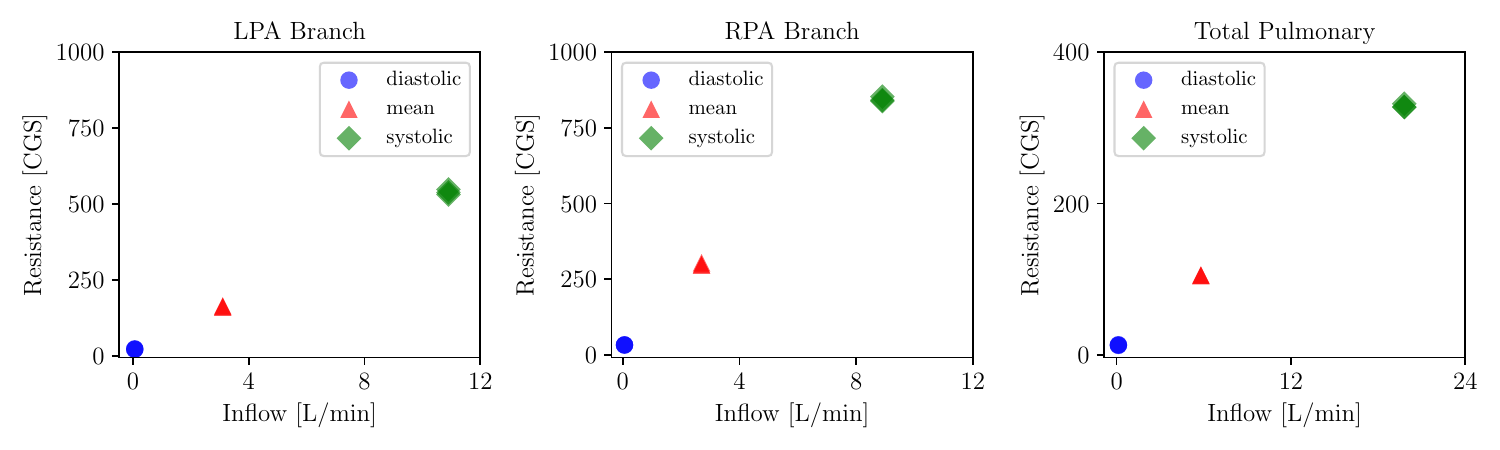}
    \caption{Relation between inflow and resistance for the entire model, left, and right pulmonary branches. See Appendix~\ref{sec:app_nonlin_res} for additional details.}\label{fig:Q_vs_R}
\end{figure}

\noindent A simplified 0D model, as shown in Figure~\ref{fig:tuning_model}, is constructed to quickly tune RCR boundary conditions. It comprises two fixed non-linear resistors and two tuneable RCR Windkessel elements connected to a constant left atrial pressure.
During the tuning process, the two non-linear resistors approximate the total pulmonary artery resistance in the left and right pulmonary artery trees.
To determine the relation between flow and pressure drop in each non-linear resistor, we ran three steady-state 3D simulations of the corresponding model at diastolic, mean, and systolic inflow values, respectively.
For each simulation, every outlet was coupled to a resistance boundary condition and approximated by splitting the total pulmonary vascular resistance (PVR) according to Murray's Law~\cite{murray1926physiological} (See Equation \ref{eq:murrays_law}). A constant zero atrial pressure was maintained.
We measured the pressure at the inlet and outlets on each side, determined the pressure drops, and averaged it over the number of outlets on each side. Different approaches used to compute this average (arithmetic, area and inverse area weighted) were found to produce similar results.
Similarly, we computed the flow split toward each branch and found the resistance value through the generalized Ohm's Law ($R = \Delta P / Q$). Repeating this procedure for each simulation and plotting flows versus resistances (see Figure~\ref{fig:Q_vs_R}), we found a linear relationship between the two values, expressed as $R = sQ$.
Substituting $sQ$ for $R$ in Ohm's Law, we obtain the equation for a quadratic resistor $RQ = sQ^2 = \Delta P$.
The slopes $s_{LPA}$ and $s_{RPA}$, for the LPA and RPA branches, respectively, were finally computed from the linear trends in Figure~\ref{fig:Q_vs_R}. For the interested reader, additional detail can be found in Appendix~\ref{sec:app_nonlin_res}.

The tuneable RCR Windkessel elements represent an aggregate of the RCR boundary conditions for both LPA and RPA outlets.
However, due to a limited set of 4 physiological targets but 6 unknown parameters, we make several assumptions to ensure identifiability. 
First, we assume a ratio of $1$ to $9$ between the proximal and distal resistances, tuning them as a single resistance $R_{capillary}$. 
Additionally, we only tune for a single capacitance $C$ shared between the two pulmonary branches. We initialize the capacitors and resistors using
\begin{equation}
 C = 10^{-3}\,\text{mL/Barye},\quad R_{LPA,\mathrm{capillary}} = R_{RPA,\mathrm{capillary}} = \text{PVR},
\end{equation}
where PVR is the pulmonary vascular resistance computed as 
\[
\text{PVR} = \frac{P_{\mathrm{mean}} - \text{PCWP}}{Q_{\mathrm{inflow}}},
\]
$P_{\mathrm{mean}}$ is the mean pulmonary artery pressure, and \text{PCWP} is the capillary wedge pressure.

To determine the final RCR values, we employ Nelder-Mead optimization~\cite{nelder1965simplex} to minimize a loss function $\mathcal{L}$ expressed as the average of four squared error contributions, where $P$ is the pulmonary artery pressure, $Q$ is flow, and $f_{\mathrm{RPA}}$ is the RPA flow split, leading to
\begin{equation}\label{equ:loss_bc_tuning}
\mathcal{L} = \left(\frac{\hat{P}_{\mathrm{sys}} - P_{\mathrm{sys}}}{P_{\mathrm{sys}}}\right)^2
+ \left(\frac{\hat{P}_{\mathrm{mean}} - P_{\mathrm{mean}}}{P_{\mathrm{mean}}}\right)^2
+ \left(\frac{\hat{P}_{\mathrm{dia}} - P_{\mathrm{dia}}}{P_{\mathrm{dia}}}\right)^2
+ \left(\frac{\hat{Q}_{\mathrm{RPA}} - f_{\mathrm{RPA}}\cdot Q_{\mathrm{inflow}}}{f_{\mathrm{RPA}}\cdot Q_{\mathrm{inflow}}}\right)^2.
\end{equation}
Once tuning was completed, the RCR values were split into individual outlets according to Murray's Law~\cite{murray1926physiological}, i.e.
\begin{equation}\label{eq:murrays_law}
R_{p,i} = \frac{A}{A_i}R_p,\quad C_i = \frac{A_i}{A}C\quad R_{d,i} = \frac{A}{A_i}R_d,
\end{equation}
where $A$ is the sum of cross-sectional areas across all the model outlets, and $A_i$ is the $i$-th outlet where a proximal resistor $R_{p,i}$, a distal resistor $R_{d,i}$ and a capacitor $C_i$ are respectively applied. 
This corresponds to assuming a parallel arrangement for the time-constant RCR circuits at each outlet.
An alternative formulation for the boundary conditions is discussed in~\cite{virtual_intervention} where the MPA inlet is coupled to a right heart lumped parameter model with constant right atrial pressure, and circuit elements represeting the tricuspid, pulmonary valve and the right ventricle.

\subsection{Automatic generation of 0D reduced order models}\label{sec:lpn}

\begin{figure}[ht!]
\centering
\begin{subfigure}[b]{0.4\linewidth}
    \includegraphics[width=\textwidth]{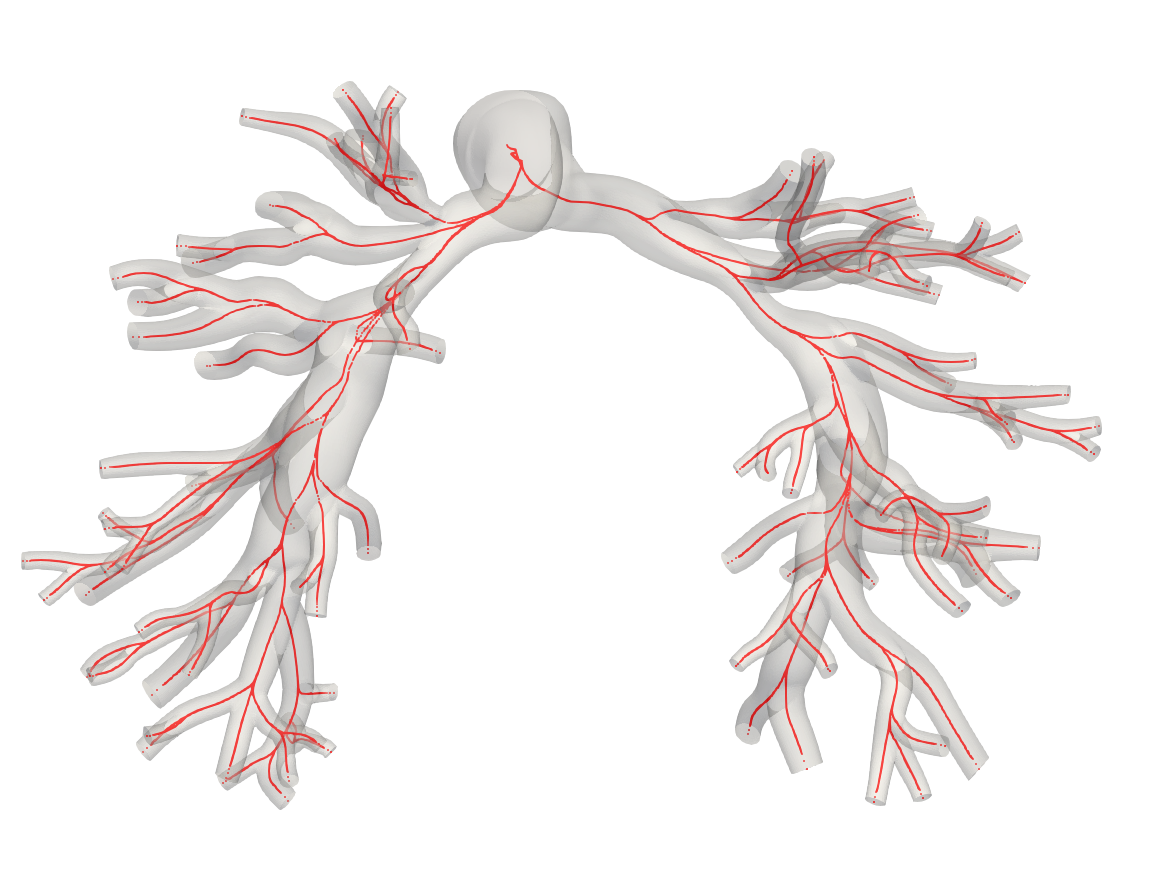}
    \caption{Extraction of centerline paths}\label{fig:centerlines}
    \label{fig:AS1_SU0308_centerlines}
\end{subfigure}
\begin{subfigure}[b]{0.4\linewidth}
    \includegraphics[width=\textwidth]{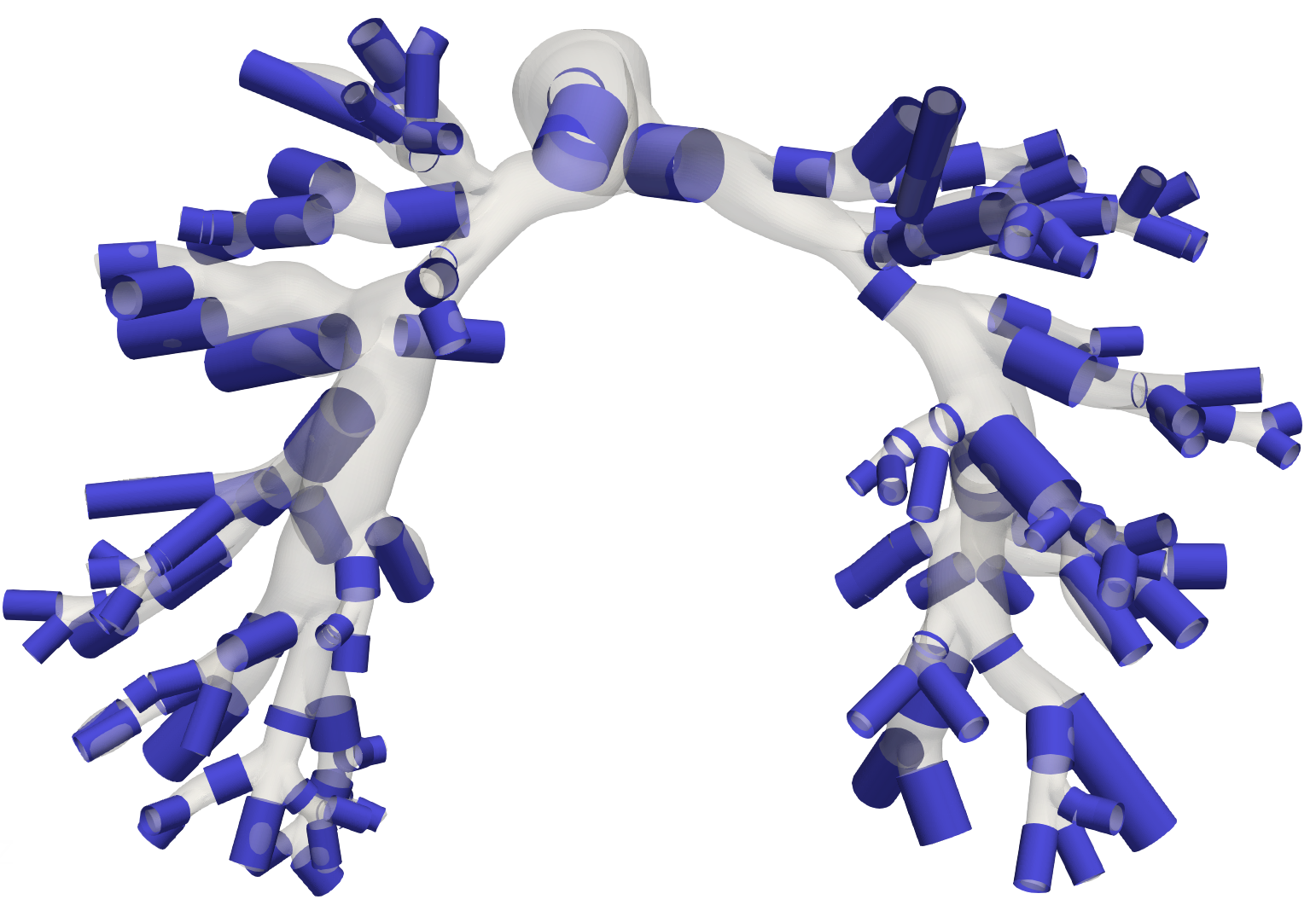}
    \caption{0D-superimposed model}
    \label{fig:modeled_region}
\end{subfigure}
\caption{A visualization of the extracted centerlines (left) and the generated 0D circuit elements superimposed on the diseased anatomy (right). The cylindrical regions in blue represent explicitly modeled circuit elements, while the semi-transparent 3D regions are treated as \emph{junctions}. Evidently, a 0D representation is unable to resolve the geometrical complexity of a 3D pulmonary model, as a significant portion of the pulmonary tree is assigned to junctions.}
\end{figure}

\noindent We utilize a recently developed automated model reduction pipeline to generate a \emph{low-fidelity} 0D model from the diseased pulmonary model~\cite{automated_0d}.
This pipeline converts the branches of a 3D geometry into one or more segments, and assigns RCL elements to each segment, computed assuming Poiseuille flow conditions.
We first begin by extracting the centerlines from our 3D diseased anatomy using Simvascular's VMTK interface~\cite{antiga2008image} (see Figure~\ref{fig:centerlines}). 
In this step, the 3D geometry is separated into branches and junctions, storing this information within the centerline representation, facilitating the association between branches and circuit elements.
From the centerlines, the number of segments per branch is then determined by an automated stenosis detection module which analyzes the change in lumen radius along the branch. If no relative extrema are found in a given branch, the module creates a single RCL element.
%
Otherwise, depending on the location and number of extrema, the branch is subdivided into either 1 or 3 RCL elements, containing an additional stenosis coefficient corresponding to a non-linear quadratic resistor.
When referring to the 0D model, each RCL element is identified as a \emph{vessel}, and a vascular branch may be comprised of multiple vessels.
For each segmented vessel, the RCL constants related to the viscous resistance, elastic vessel compliance, and blood inertia are determined according to the following expressions (see, e.g.~\cite{milivsic2004analysis}).
\begin{equation}
R = \frac{8{\mu}l}{{\pi}r^4},\quad C = \frac{3l{\pi}r^3}{2Eh},\quad L=\frac{{\rho}l}{{\pi}r^2},
\end{equation}
where $\mu$ is blood viscosity assumed equal to 0.04 poise, $\rho$ is blood density set equal to 1.06 g$\cdot$cm$^-3$, and $l$ is the length of the selected vessel segment. 
Additionally, $r$ is the average radius of the vessel, since we model each vessel with an ideal cylindrical lumen, $E$ is the modulus of elasticity assuming an isotropic and homogeneous elastic vessel response, and $h$ is the vessel thickness.  
In this study, we assume a linear ratio $Eh/r=2.5 * 10^6$ dynes/cm$^2$ consistent with~\cite{virtual_intervention}.
Lastly, the additional stenosis coefficient is defined by 
\begin{equation}
S = K_t\cdot\frac{\rho}{2S_0^2}\cdot\left(\frac{S_0}{S_s} - 1\right)^2,
\end{equation}
where $S_0$ and $S_s$ represent the cross-sectional lumen areas proximal to and at the stenosis, respectively, and $K_t$ is a commonly used empirical correction factor set at $1.52$.

Finally, conservation of mass and pressure continuity are enforced at the junctions. 
Boundary conditions remain identical to 3D formulations, with a prescribed inlet waveform, RCR Windkessel elements coupled to each outlet, and a constant left atrial pressure.
All 0D lumped parameter models are solved using svZeroDPlus, a C++ implementation of the Python-based svZeroDSolver available at \url{https://github.com/StanfordCBCL/svZeroDPlus}. All simulations were run for 6 cardiac cycles to ensure convergence.

\subsection{Matching 3D and 0D hemodynamics by iterative linear corrections}\label{sec:linear_correction}

\noindent Lumped parameter hemodynamic models are based on the same equations governing the evolution of current and voltage in an electrical circuit. At junctions, these equations typically assume a unique value of the pressure and flow continuity, failing to account for minor losses at bifurcations.
Although this simplification may have negligible effects for some anatomies (see, e.g.,~\cite{automated_0d}), the large number of branches in pulmonary models exasperates this issue, leading to significant differences in the resulting pressure distribution (e.g., $\sim$10\% deviation from three-dimensional simulations as shown in~\cite{automated_0d}). 
A visual representation of the geometrical complexity of regions in the 3D model treated as junctions is offered in Figure~\ref{fig:modeled_region}, suggesting the need to introduce some correction term to mitigate the 3D to 0D model discrepancy.

To measure this discrepancy,  we integrated the three-dimensional model pressure and flow values over a number of cross-sectional slices and projected the results onto the centerlines. We then projected the 0D solution at the input and output of each 0D vessel onto identical centerlines.

\begin{figure}[ht!]\newcommand\tw{0.065}
\centering
\begin{subfigure}[b]{\linewidth}
    \includegraphics[width=\textwidth]{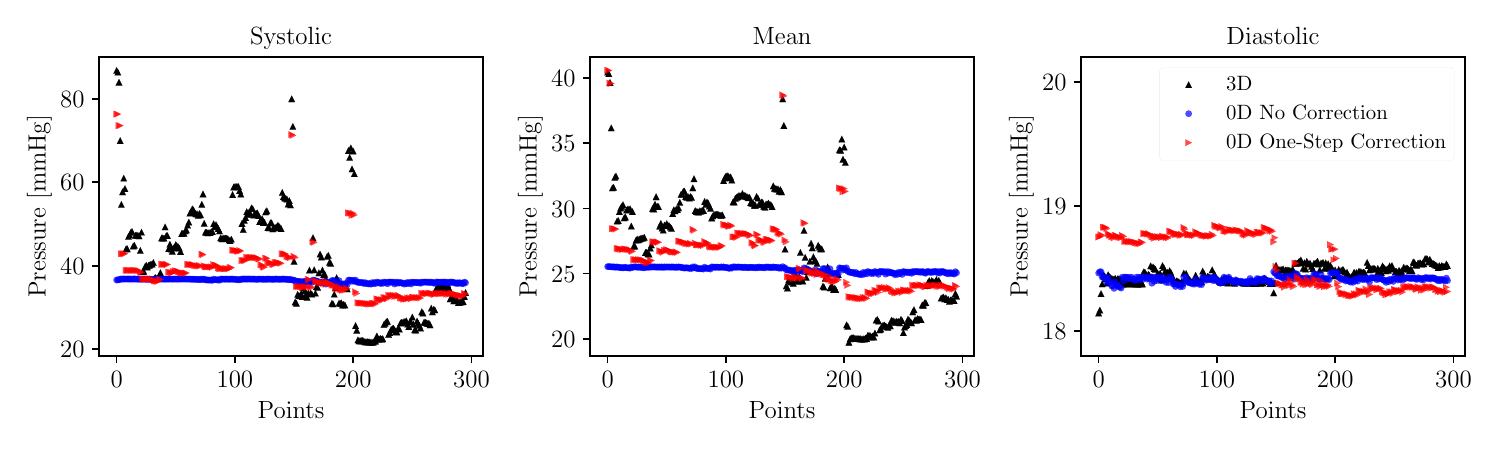}
    \caption{3D and 0D pressure output along the centerline for the pre-operative configuration.}
    \label{fig:original_0D_1}
\end{subfigure}
\begin{subfigure}[b]{\linewidth}
    \includegraphics[width=\textwidth]{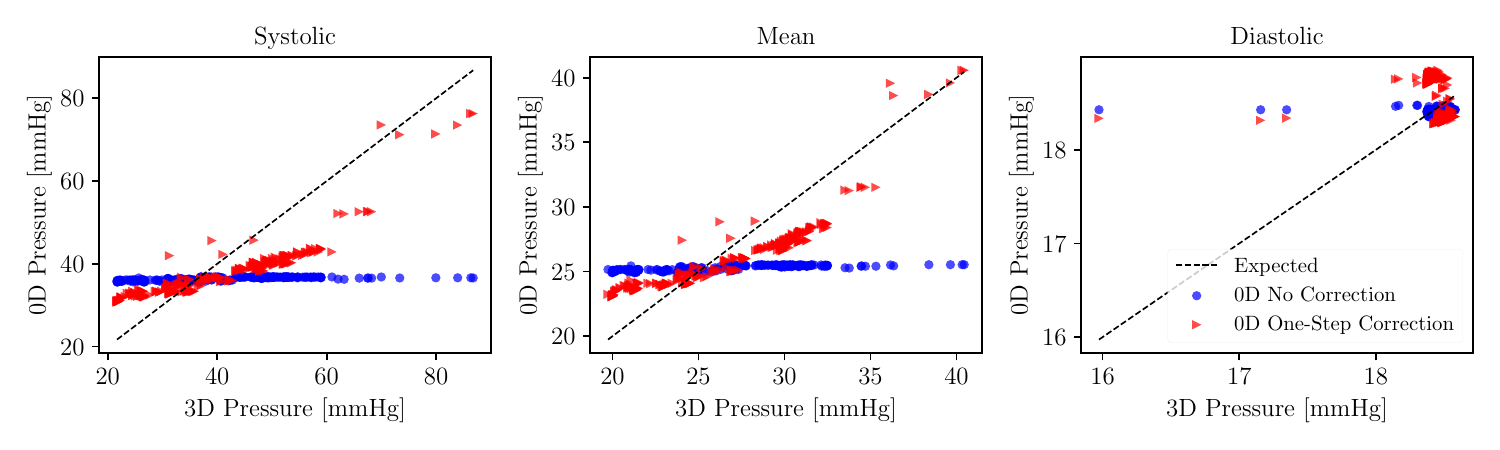}
    \caption{Agreement between 3D and 0D pressure output for the preoperative configuration.}
    \label{fig:original_0D_2}
\end{subfigure}
\caption{A comparison between 3D, uncorrected 0D and one-step linear corrected 0D models. The uncorrected 0D model (blue) captures a negligible pressure drop and incorrect distribution across the pulmonary tree. Our one-step correction (red) results in improved agreement with the 3D result.}
\label{fig:original_0D}
\end{figure}

We then introduced a simple yet effective approach to improving the agreement between zero- and three-dimensional model outputs, which consists of adding a linear resistor at the outlet of every junction and performing a linear correction. 
Consider $d$ locations in the pulmonary artery tree, where each location is a newly initialized resistor ($R = 0$) at the junction outlets. 
Also, consider a collection of unitary resistance increments $(\boldsymbol{\Delta R}_{1},\boldsymbol{\Delta R}_{2},\dots,\boldsymbol{\Delta R}_{d})$ applied to the corresponding locations, i.e., $\boldsymbol{\Delta R}_i$ has a $i$-th component equal to $1$ and zero for all the other components.
When solving the 0D model with a unit resistance change at the $i$-th location $\boldsymbol{\Delta R}_{i}$, a corresponding change in mean pressure $\Delta P_{j,i}$ is observed at an arbitrary location $j$. 
Therefore, if we assume \emph{linear superposition} of effects, a change in resistance equal to 
\[
\boldsymbol{\Delta R} = \alpha_{1}\,\boldsymbol{\Delta R}_{1} + \alpha_{2}\,\boldsymbol{\Delta R}_{2},\dots,\alpha_{d}\,\boldsymbol{\Delta R}_{d},
\]
would lead to a total change in mean pressure at location $j$ equal to 
\[
\Delta P_{j} = \sum_{i=1}^{d}\,\Delta P_{j,i}\,\alpha_{i} + \beta,
\]
where $\beta$ accounts for a bias term representing a spatially uniform change in pressure that is independent on $\boldsymbol{\Delta R}$.
Now we determine the coefficients $\alpha_{i},i=1,\dots,d$ and $\beta$ so that the total change in pressure at location $j$, $\Delta P_{j},\,j=1,\dots,l$ corresponds to the difference between mean 3D and 0D model pressures, i.e. $P^{3D}_{j} - P^{0D}_{j}$, leading to the system 
\begin{equation}\label{equ:system}
\sum_{i=1}^{d}\,\Delta P_{j,i}\,\alpha_{i} + \beta = P^{3D}_{j} - P^{0D}_{j},\,\,\text{for}\,\,j=1,\dots,l,
\end{equation}
or in expanded form
\begin{equation}
\begin{bmatrix}
\Delta P_{1,1} & \Delta P_{1,2} & \dots & \Delta P_{1,d} & 1\\
\Delta P_{2,1} & \Delta P_{2,2} & \dots & \Delta P_{2,d} & 1\\
\vdots & \vdots & \vdots & \vdots\\
\Delta P_{l,1} & \Delta P_{l,2} & \dots & \Delta P_{l,d} & 1\\
\end{bmatrix}\cdot
\begin{bmatrix}
\alpha_{1}\\
\alpha_{2}\\
\vdots\\
\alpha_{d}\\
\beta\\
\end{bmatrix} = 
\begin{bmatrix}
\Delta P^{*}_{1}\\
\Delta P^{*}_{2}\\
\vdots\\
\Delta P^{*}_{l}\\
\end{bmatrix},
\text{where}\,\,\Delta P_j^* = P^{3D}_{j} - P^{0D}_{j},\,\text{and } \, l = d + 1.
\end{equation}

By solving for the coefficients $\boldsymbol{\alpha}$, we determine the resistance values at all the bifurcation outlets. Additionally, the coefficient $\beta$ suggests a constant pressure gap that can be applied by changing the total boundary resistance. This is obtained using Murray's Law, maintaining the existing ratio of proximal to distal resistance.
Nonetheless, since this method strictly defines a mathematical relationship, $\alpha$ may occasionally be less than zero, leading to negative resistances that contradict the physical principles governing our equations. To address this issue, we disregard any negative values and uphold the original zero resistance. 
Figure~\ref{fig:original_0D} shows the result of applying the proposed linear correction.

In spite of its simplicity, the proposed approach is completely non-intrusive and can be generalized in several ways. For example, the locations for the $d$ additional resistances and $l$ measurements can be selected arbitrarily, as long as the system~\eqref{equ:system} can be solved. 
While our correction is formulated in terms of pressures due to the main difference between 3D and 0D models being the minor loss at bifurcations, additional rows can be added to~\eqref{equ:system} to account for discrepancies in the flow rate too. 
Also, changes in capacitance and inductance can also be easily added to the proposed approach. 
Other enhancements relate to selective applications to model subdomains, changes in the resistance of existing vessels, and performing multiple iterations. These latter scenarios are discussed in the next sections. 

\subsubsection{Sub-domain corrections}\label{sec:subdomain}

\begin{figure}[ht!]\newcommand\tw{0.065}
\centering
\begin{subfigure}[b]{\linewidth}
    \includegraphics[width=\textwidth]{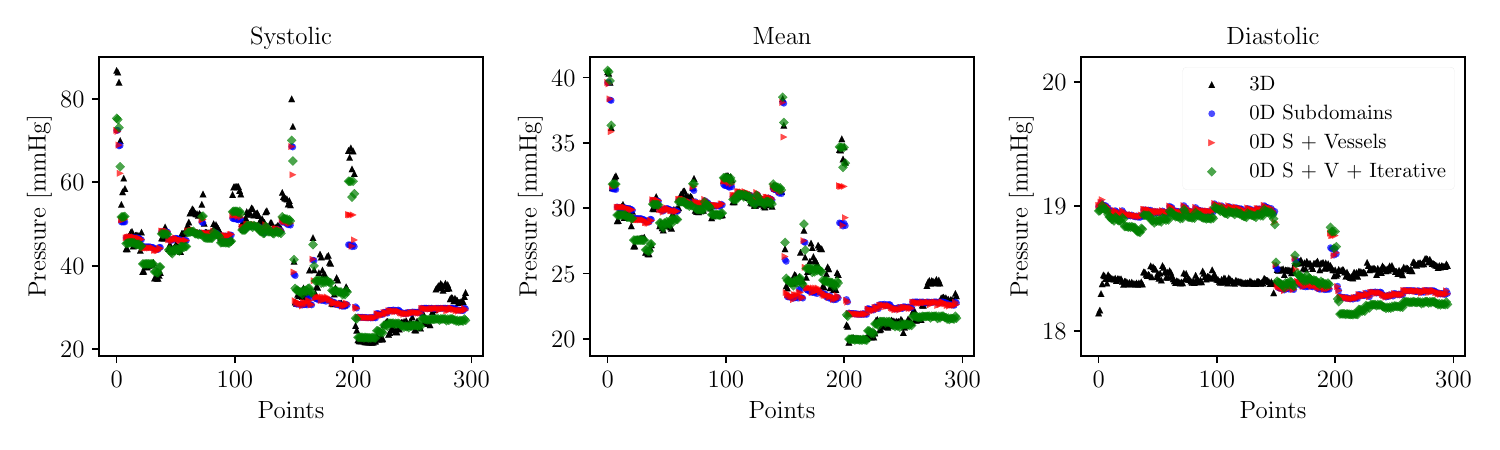}
    \caption{3D and 0D pressure output along the centerline for extensions to the linear correction method.}
    \label{fig:final_lc_1}
\end{subfigure}
\begin{subfigure}[b]{\linewidth}
    \includegraphics[width=\textwidth]{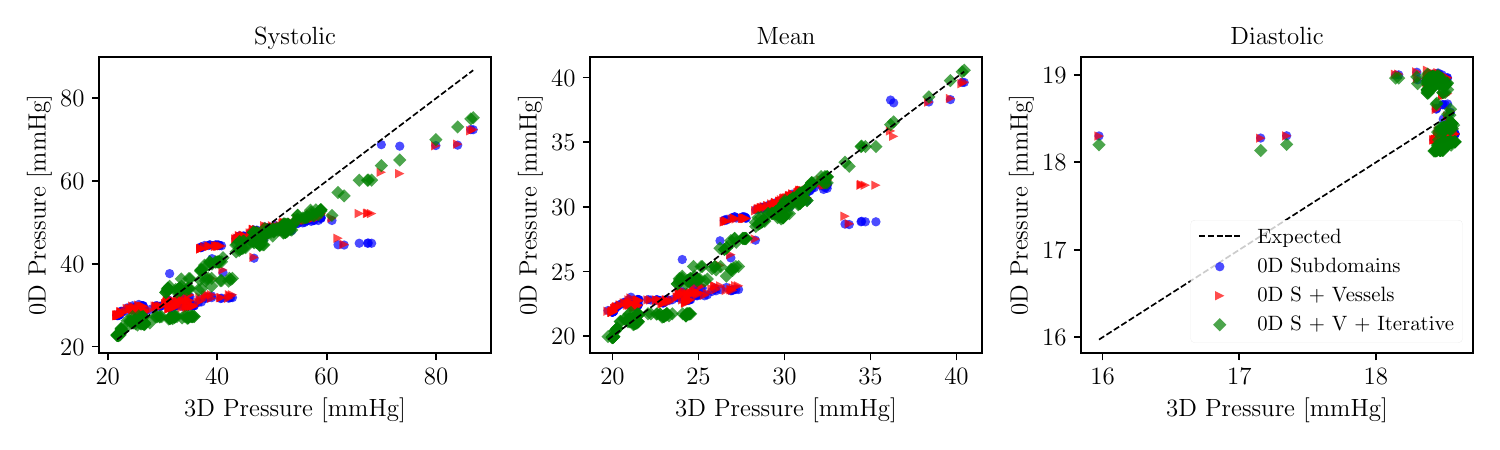}
    \caption{Agreement between 3D and extended corrected 0D pressure outputs.}
    \label{fig:final_lc_2}
\end{subfigure}
\caption{A comparison between 3D pressure values and the linear correction with progressively more enhancements: subdomains, vessels, and iterative corrections. The final version of linear correction (green) demonstrates a much stronger match to the 3D truth.}
\label{fig:final_lc}
\end{figure}

\noindent Figure~\ref{fig:original_0D_1} shows marked differences between pressures in the left and right pulmonary arteries (LPA and RPA, respectively). 
Hypothesizing a better correction when treating certain regions in isolation, we experimented with three sequential corrections in the MPA, RPA, and LPA. 
The MPA includes the initial inlet branch and the first junction that splits into the RPA and LPA. All the downstream vessels of the first junction are then grouped into the RPA or LPA. 
For each correction, although the set of $d$ locations is reduced, we continue to include $\Delta P^*_{\mathrm{MPA}}$ and bias $\beta$ to account for spatially uniform pressure corrections.
After each step, $\alpha$ and $\beta$ values were applied such that the next step included prior changes. 
The results of this procedure, as shown in Figure~\ref{fig:final_lc}), suggest that partitioning does indeed achieve closer matches. In fact, even finer partitions may prove to be beneficial.
Finally, changing the order of the correction (MPA first, followed by LPA and RPA) did not significantly affect the final result. 
\subsubsection{Correction of existing vessel resistance}

\noindent Upon review, it was observed that corrected 0D models only captured the average 3D model pressures at each subdomain without accurately representing local pressure variability.
While we had previously calibrated only new resistances at each junction outlet, we generalized our approach by also calibrating the resistances at all vessels, represented through RCL elements.
These vessel corrections have a marginal effect on the pressure distribution (Figure~\ref{fig:final_lc}).
Nonetheless, given the enhanced accuracy at certain locations, we decided to maintain vessel elements as part of our correction methodology.

\subsubsection{Iterative Linear Correction}

\noindent Finally, to achieve a stable correction given the non-linearity introduced by zeroing out negative resistances, we chose to repeat the sequential correction multiple times. 
After empirical testing, we found that five iterations were sufficient for convergence. The results in Figure~\ref{fig:final_lc} demonstrate excellent agreement between the 3D model and corrected 0D model solutions. This is confirmed by the significantly higher absolute Pearsons correlations shown in Table~\ref{table2} for the systolic and mean pressures. Diastolic pressures are instead spatially uniform through the pulmonary arterial tree and well approximated even with no correction.

\begin{figure}[ht!]\newcommand\tw{0.065}
\centering
\begin{subfigure}[b]{.28\linewidth}
    \includegraphics[width=\textwidth]{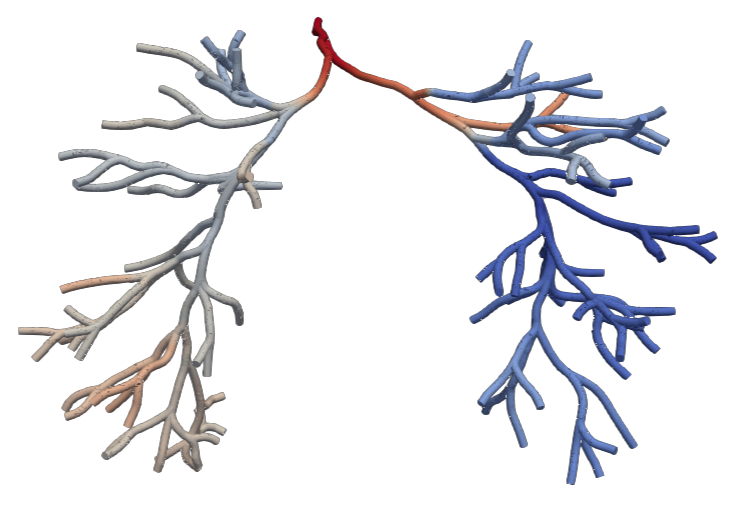}
    \caption{3D solutions}
\end{subfigure}
\begin{subfigure}[b]{.28\linewidth}
    \includegraphics[width=\textwidth]{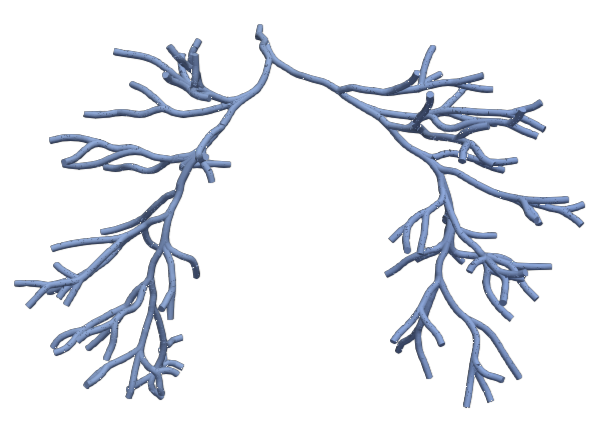}
    \caption{0D solution w/o correction}
\end{subfigure}
\begin{subfigure}[b]{.28\linewidth}
    \includegraphics[width=\textwidth]{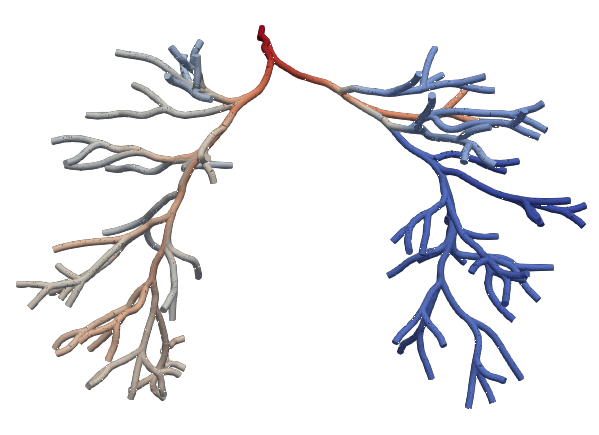}
    \caption{0D solution w/ correction}
\end{subfigure}
\begin{subfigure}[b]{.095\linewidth}
    \includegraphics[width=\textwidth]{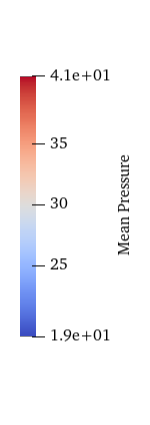}
\end{subfigure}
\centering
\begin{subfigure}[b]{.4\linewidth}
    \includegraphics[width=\textwidth]{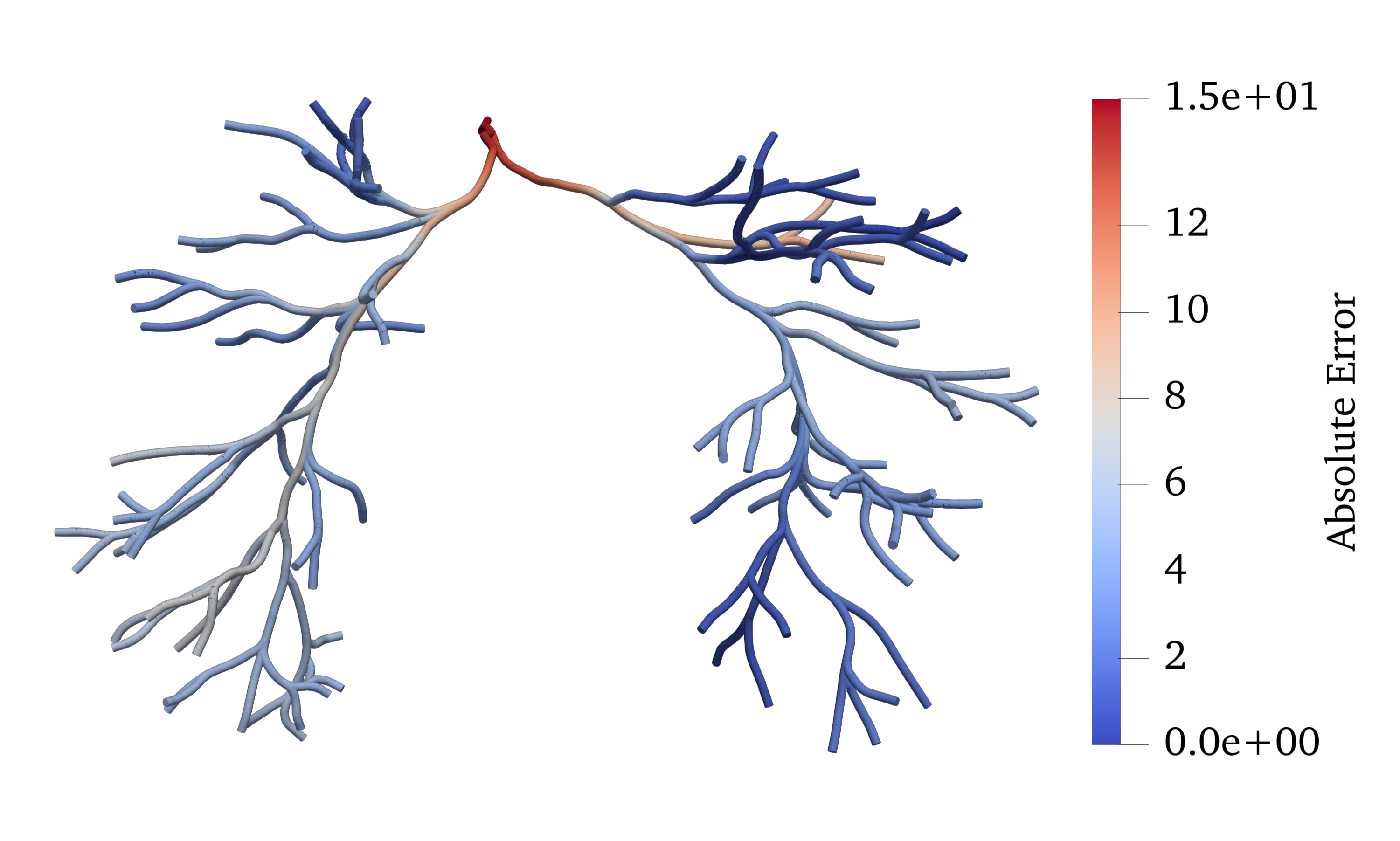}
    \caption{Absolute Error between 3D and uncorrected 0D solution}
\end{subfigure}
\begin{subfigure}[b]{.4\linewidth}
    \includegraphics[width=\textwidth]{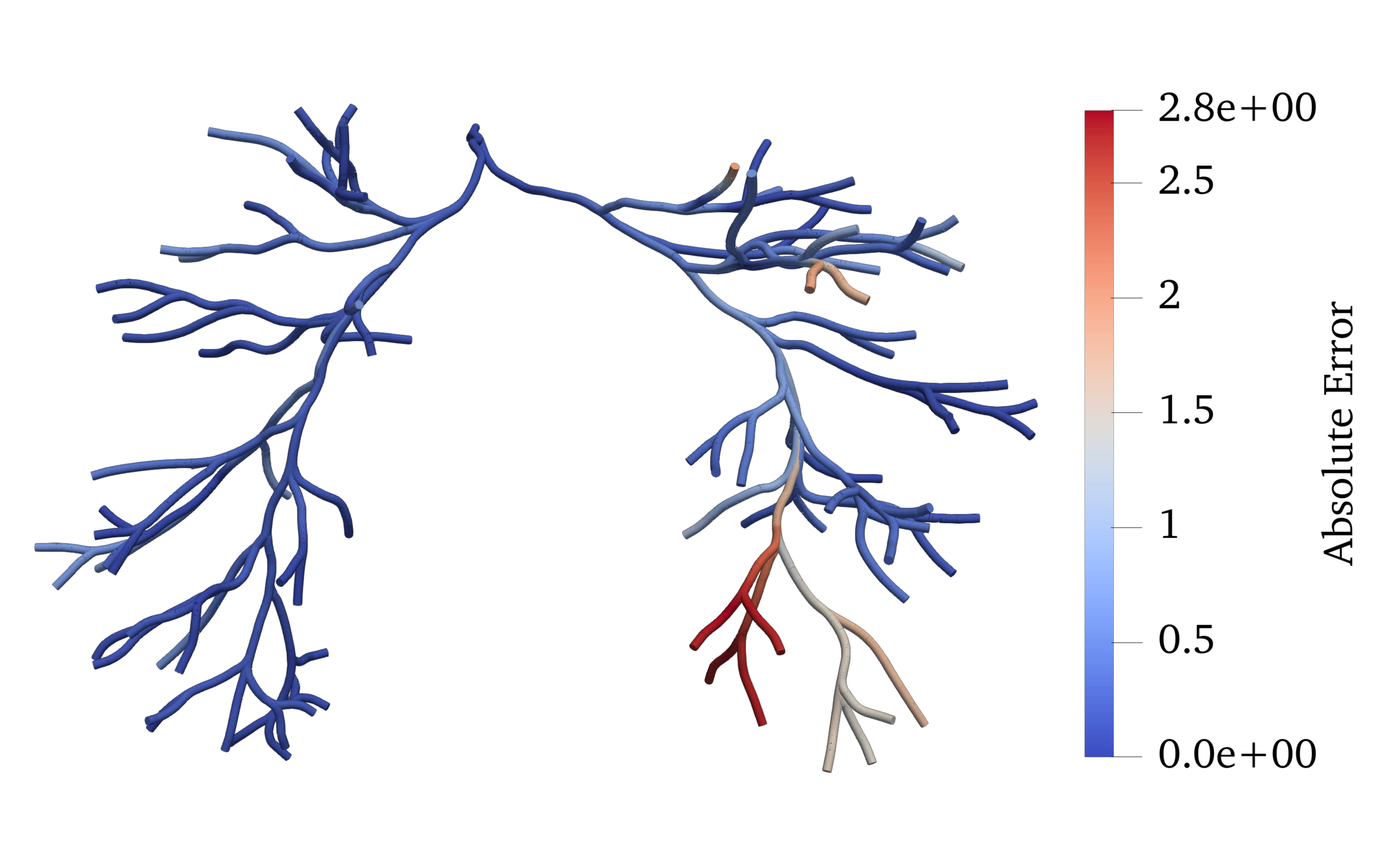}
    \caption{Absolute Error between 3D and final corrected 0D solution}
\end{subfigure}
\caption{A visual comparison of mean pressure distribution for the 3D solution (a), non-corrected 0D solution (b), and final corrected 0D solution (c). Absolute errors of $|\mathrm{(a)} - \mathrm{(b)}|$ and $|\mathrm{(a)} - \mathrm{(c)}|$ are shown in (d) and (e) respectively (with different scales). Since 0D solutions are found at discretized points along the centerlines, a linear interpolation of pressures and errors is used to generate the continuous measurements along the centerlines in images (b-d). We note that the localized region of higher error even within our improved 0D formulation may be mitigated by selecting smaller subdomains, as suggested in~\ref{sec:subdomain}.   }
\label{fig:centerline_3d_0d}
\end{figure}

Figure~\ref{fig:centerline_3d_0d} shows a direct comparison of hemodynamic results along the model centerline, demonstrating the limited accuracy of pressure drops and distributions initially achieved through the 0D model with optimally tuned boundary conditions. 

\begin{table}[ht!]
\centering
\caption{Absolute value of the Pearson correlation coefficient for each method mentioned in Figure~\ref{fig:original_0D} and~\ref{fig:final_lc} when compared to 3D solutions.}
\begin{tabular}{l c c c c c}
\toprule
{\bf} & {\bf No Correction} & {\bf One-Step} & {\bf Subdomains} & {\bf S + Vessels} & {\bf S + V + Iterative }\\
\midrule
{\bf Systolic} & 0.743 & 0.880  & 0.925 & 0.941 & 0.977\\
{\bf Mean} & 0.831 & 0.890  & 0.941 & 0.954 & 0.987\\
{\bf Diastolic} & 0.051 & 0.138 & 0.147 & 0.145 & 0.118\\
\bottomrule

\end{tabular}
\label{table2}
\end{table}

\section{Parameterizing Repairs}\label{sec:parameterization}
%
\noindent The correction introduced in Section~\ref{sec:linear_correction} modifies the resistance distribution in the 0D model, significantly improving the agreement with 3D model results. 
Nonetheless, the same correction can be used to find a set of resistance increments representing a \emph{transition} between two distinct 3D model configurations and corresponding results, in our case a \emph{diseased} and \emph{repaired} (or pre- and post-operative) configurations.
Consider two resistance distributions $\boldsymbol{\Delta R}_{i,o\rightarrow d}$, $\boldsymbol{\Delta R}_{i,d\rightarrow r}$ applied to the 0D model through the proposed correction, where $\boldsymbol{\Delta R}_{i,o\rightarrow d}$ is the correction from an \emph{original} to a \emph{diseased} state and $\boldsymbol{\Delta R}_{i,d\rightarrow r}$ is the correction from a \emph{diseased} to a \emph{repaired} state following treatment at the $i$-th stenotic lesion. 
Then a possible parametrization between these two configurations would be 
\begin{equation}\label{equ:0D_param}
\boldsymbol{\Delta R_i}(c_{i}) = 
\boldsymbol{\Delta R}_{i,o\rightarrow d} + (c_{i})\boldsymbol{\Delta R}_{i,d\rightarrow r}.
\end{equation}
In the next section, we discuss how we generated repaired configurations for the three-dimensional anatomy and how the parameterization in~\eqref{equ:0D_param} was implemented. 

\subsection{Three-dimensional anatomical repairs}\label{sec:repair_3D}
\begin{figure}[ht!]
\centering

\begin{subfigure}[b]{.47\linewidth}
    \includegraphics[width=\textwidth]{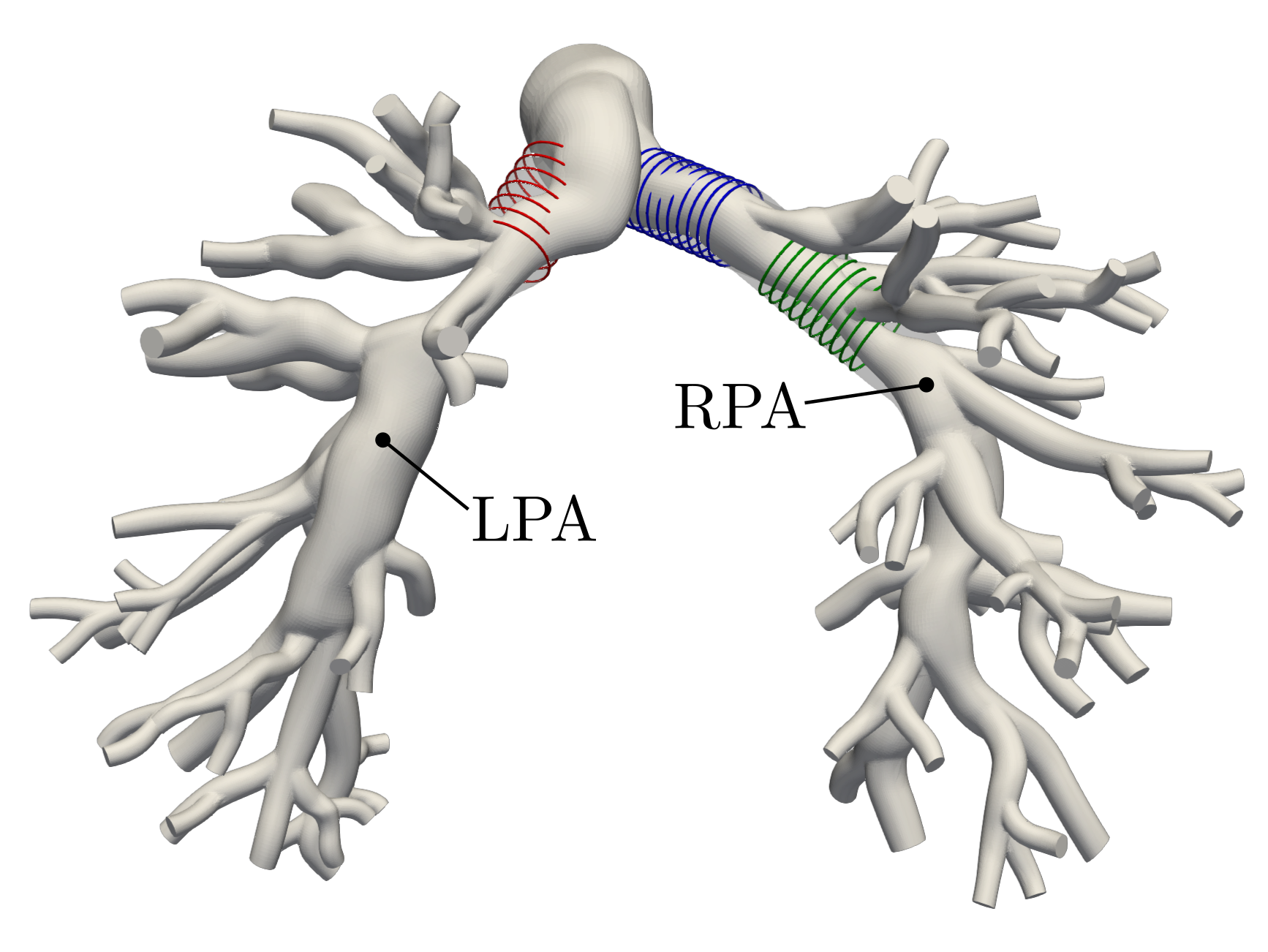}
    \caption{Repaired 3D Segmentations.}
    \label{fig:repaired_3d}
\end{subfigure}
\begin{subfigure}[b]{.47\linewidth}
    \includegraphics[width=\textwidth]{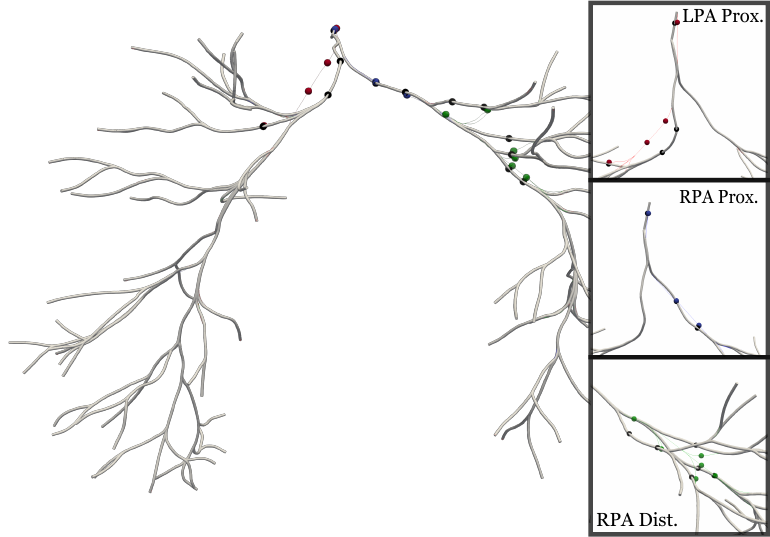}
    \caption{Repaired centerline matching.}
    \label{fig:repaired_centerlines}
\end{subfigure}

\caption{The repaired segmentations (a) and close-up centerline matching (b). The repairs are denoted LPA proximal (red), RPA proximal (blue), and RPA distal (green). For brevity, repaired segmentations and centerlines are compressed into a single image but exist in practice as three separate models. The centerline matching indicates how relevant measurement points are mapped from repaired centerlines (colored) to the original diseased centerline (black) in order to preserve consistency.}
\label{fig:repaired_model_cent}
\end{figure}

%
\noindent For the selected diseased model (AS1\_SU0308), three stenotic lesions amenable to repair were identified, and three virtually stented configurations were segmented under the supervision of a clinician, see~\cite{virtual_intervention}.
In what follows, we refer to the locations of these lesions as the \emph{proximal LPA}, \emph{proximal RPA}, and \emph{distal RPA} repair, respectively. 
For each configuration, the region undergoing repair was segmented using the expected post-repair geometry and lofted into the 3D geometries visualized in Figure~\ref{fig:repaired_3d}. 
We computed hemodynamic results for these three repaired models as discussed in Section~\ref{sec:threed_modeling}, using the same boundary conditions as the diseased model, then extracted the centerlines, integrated 3D pressures and flows over several cross-sections, and finally, projected the results onto the extracted centerlines.
However, virtual stenting led to extracted repaired centerlines differing from the original pre-operative centerlines in the repair region. 
To maintain consistency, we mapped hemodynamic results from the repaired centerlines onto their closest locations along the diseased model centerline for each branch, disregarding junctions. 
The relevant centerline mapping for AS1\_SU0308 is illustrated in Figure~\ref{fig:repaired_centerlines}.

\subsection{Representing repairs through corrected 0D models}\label{sec:repair_0D}

\noindent Solving a 3D model for the $i$-th repaired lesion provides target hemodynamic results corresponding to $c_i = 1$, i.e., following virtual treatment.
Similarly, the original diseased zero-dimensional model serves as $c_i = 0$, where no treatment has occurred, or equivalently, a completely ineffectual repair.
Finally, we combine the repairs~\eqref{equ:0D_param} from various lesions as $\sum_{i=1}^{n} c_i\bm{\Delta R}_{i,d\rightarrow r}$, and add it to our diseased 0D model. A note regarding the limitation of this approach is mentioned in Section~\ref{sec:discussion}.
 
While, in principle, all resistances in the 0D model can be corrected to represent the hemodynamic effects of augmenting a single stenotic location, a \emph{local} change in resistance is more consistent with our independence assumption for $\rho(\bm{c})$ in~\eqref{equ:indep_c}. 
Thus, we correct the resistance only for vessel segments and junctions which correspond to the centerline regions $\mathcal{R}_{r}$ that have changed due to treatment.
To identify these regions we first computed the minimum Euclidean distance between relevant points on the diseased and repaired centerlines. In the majority of the vasculature unimpacted by repair, these distances were found to be negligible, but they were large near each stenosis. 
We then ordered centerline locations by decreasing distance and added to $\mathcal{R}_{r}$ all locations whose distance was larger than a manually defined cutoff.
We chose to retain the global bias $\beta$ while formulating the local correction, but we did not distribute the additional resistance $R_{global}$ to the boundary conditions. 
Empirically, we noticed no significant difference between including or excluding the bias, but also speculate that the bias could serve as an approximation for microvascular adaptation post-treatment.

\begin{figure}[ht!]
\centering
\includegraphics[width=\textwidth]{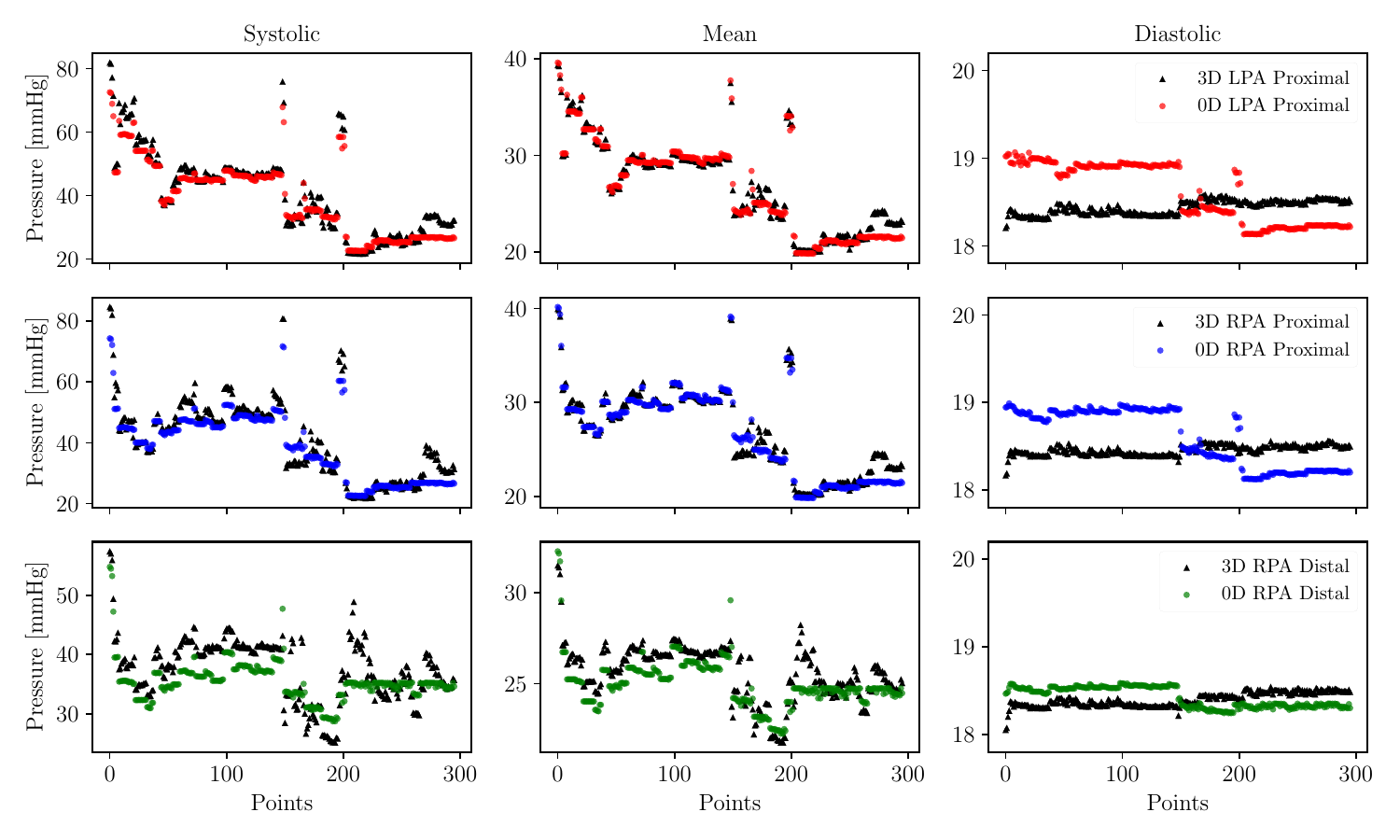}
\caption{3D model results versus \emph{locally} corrected 0D model results at the three selected stenosis locations.}
\label{fig:local_repair}
\end{figure}

Despite significantly reducing the degrees of freedom for correction, our results in Figure~\ref{fig:local_repair} continued to produce accurate results. 
However, for the distal RPA repair, the adjustment of resistances only within the RPA causes greater pressure discrepancy in the LPA, particularly at systole.

\section{Neural twin for interventional pulmonary artery rehabilitation}\label{sec:nn_surrogate}
%
\noindent Now that $p(\bm{p},\bm{f}|\bm{c})$ can be computed for any realization of the repair vector $\bm{c}$ using 0D models, we describe the neural network component of our pipeline, which is an essential component enabling the real-time analysis of interventional scenarios.
In this section, we discuss data preparation, neural network architecture, training procedure, and present the resulting model accuracy.

\subsection{Generation of Training Data}\label{sec:training_gen}

\noindent To generate sufficient training data for our neural network, we use Sobol' sampling~\cite{sobol1967distribution} to generate $32768$ ($2^{15}$) configurations of $\bm{c}$ for our training set, and $4096$ ($2^{12}$) configurations for our validation and test sets. 
To account for the varying scales across inputs and outputs, data were standardized.
For each solution $g(\bm{c})$, we extracted diastolic, systolic, and mean pressure and flow values at the main pulmonary inlet and the outlet of each vessel element (159 locations in total), amounting to 954 target values per configuration. 
These serve as our vectors $\bm{p}$ and $\bm{f}$ described in Section~\ref{sec:prob}.
While every possible measurement location provided by our 0D solver was selected in this study, a small representative set chosen by a clinician should suffice in practice.
In total, data generation took roughly 8 hours ($\sim 0.625$ seconds per simulation with postprocessing) using 24-core multiprocessing on an AMD EPYC 7543 processor with 2.8 GHz base CPU clock time and 8 GB of memory.

\subsection{Neural Network Architecture and Training}\label{sec:ann}

\noindent We then trained a feed-forward fully connected neural emulator to replace $g(\bm{c})$. 
Through empirical testing, we found the optimal architecture to consist of $|\bm{c}|$ input nodes, three hidden layers with 1000 neurons each, \emph{tanh} activations, and a linear output layer with $|\bm{p}|+|\bm{f}|$ output nodes. 
The network was trained using the Adam optimizer, with $\beta_{1} = .9$, $\beta_{2} = .999$, and an initial learning rate of $\gamma = 1e-3$ and a mean squared error (MSE) loss.
A \emph{lr-reduce-on-plateau} learning rate scheduler was also used, lowering $\gamma$ by an order of magnitude when the validation loss failed to decrease for 5 epochs, down to a minimum of $\gamma = 1e-8$. 
The training was then concluded early if the validation loss failed to decrease for another 10 epochs, or an upper limit of 500 epochs was reached. 
In most cases, the upper epoch limit was reached before early stopping conditions were met. 
Finally, the model was saved at the epoch corresponding to the lowest validation loss. For the full dataset of 32768 samples, on a single Nvidia RTX 3080Ti, training took under 20 minutes to complete.
\begin{figure}[ht!]
\centering
\includegraphics[width=.95\textwidth]{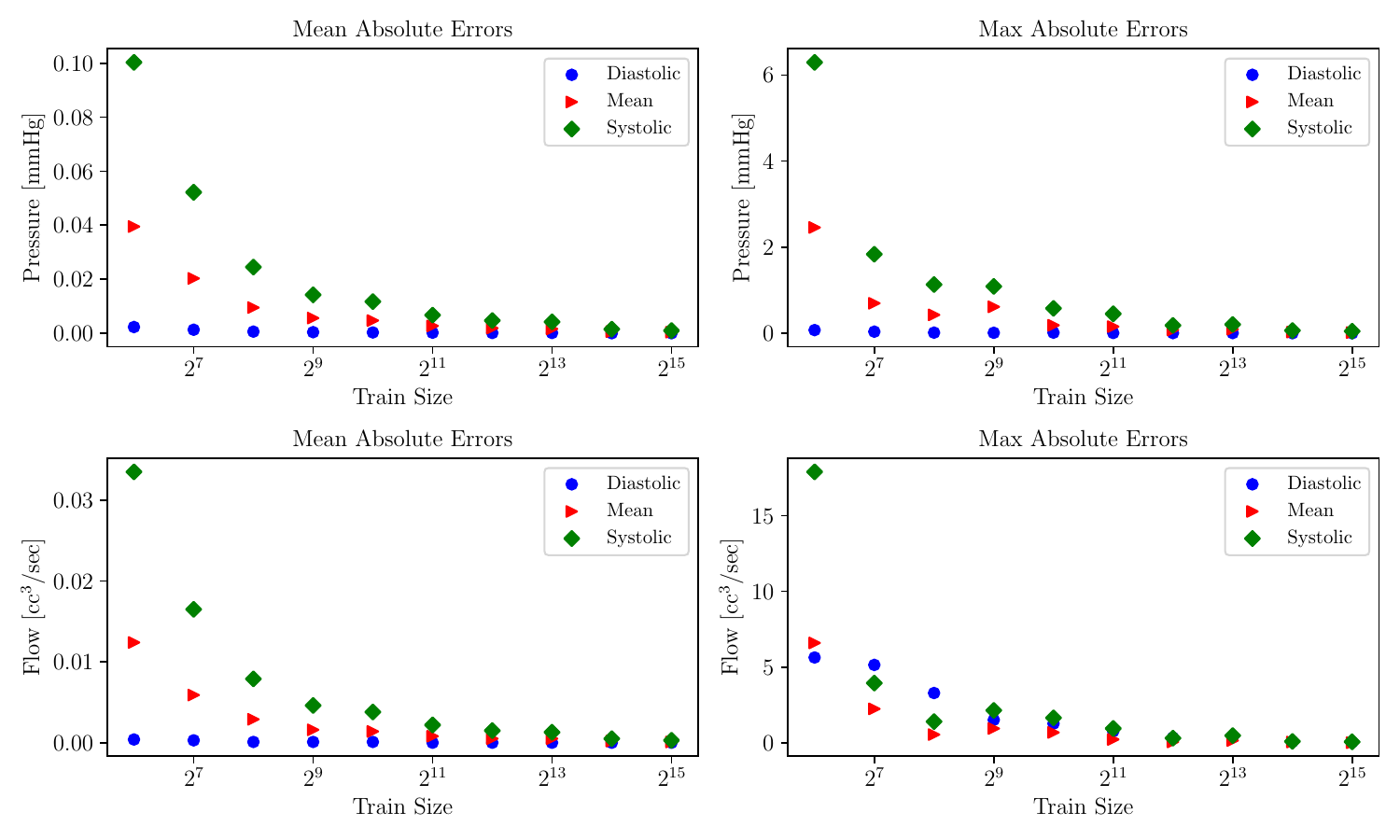}
\caption{Mean and maximum absolute errors for models trained on exponentially increasing datasets. Error measurements were taken over all diastolic, mean, and systolic pressure and flow model outputs.}\label{fig:ml_results}
\end{figure}

\subsection{Model Accuracy}\label{sec:model_acc}

\noindent While the computational cost of solving a zero-dimensional model is almost negligible, the evaluation of thousands of samples still accrues significant time. 
Thus, we evaluate the absolute flow and pressure errors of our model trained on an exponentially increasing dataset size, and plot the relationship between training data size and model accuracy in Figure~\ref{fig:ml_results}.
We noted two important conclusions from this analysis. First, a 50\% reduction in error roughly corresponds to a two-fold increase in the size of the dataset.
This provides a useful criterion to determine the optimal dataset size once an initial accuracy has been determined from a pilot training with a limited dataset, and knowing a target accuracy.
Second, this analysis shows that it is computationally feasible to train neural emulators resulting in negligible approximation error when compared to catheter measurement uncertainty.

\section{Results}\label{sec:results}

\noindent In this section, we use the proposed probabilsitic neural twin to investigate a number of treatment scenarions. 
Specifically, we consider the drastically simplified perspective of a clinician looking solely to minimize pulmonary artery pressure (PAP) and focus on addressing the following three questions:
\begin{enumerate}\itemsep 0pt
    \item[1.] What is the probability of reaching a post-operative pressure/flow within the $[p_l,p_u]$, $[f_l,f_u]$ ranges?
    \item[2.] Which repair locations are the most important to achieving the desired outcome? Additionally, how much do we need to repair in order to achieve the desired outcome?
    \item[3.] What is the optimal order for repair?
\end{enumerate}
Although we touch upon more complex considerations, we leave them to future research.

\begin{figure}[ht!]
\centering
\includegraphics[width=\textwidth]{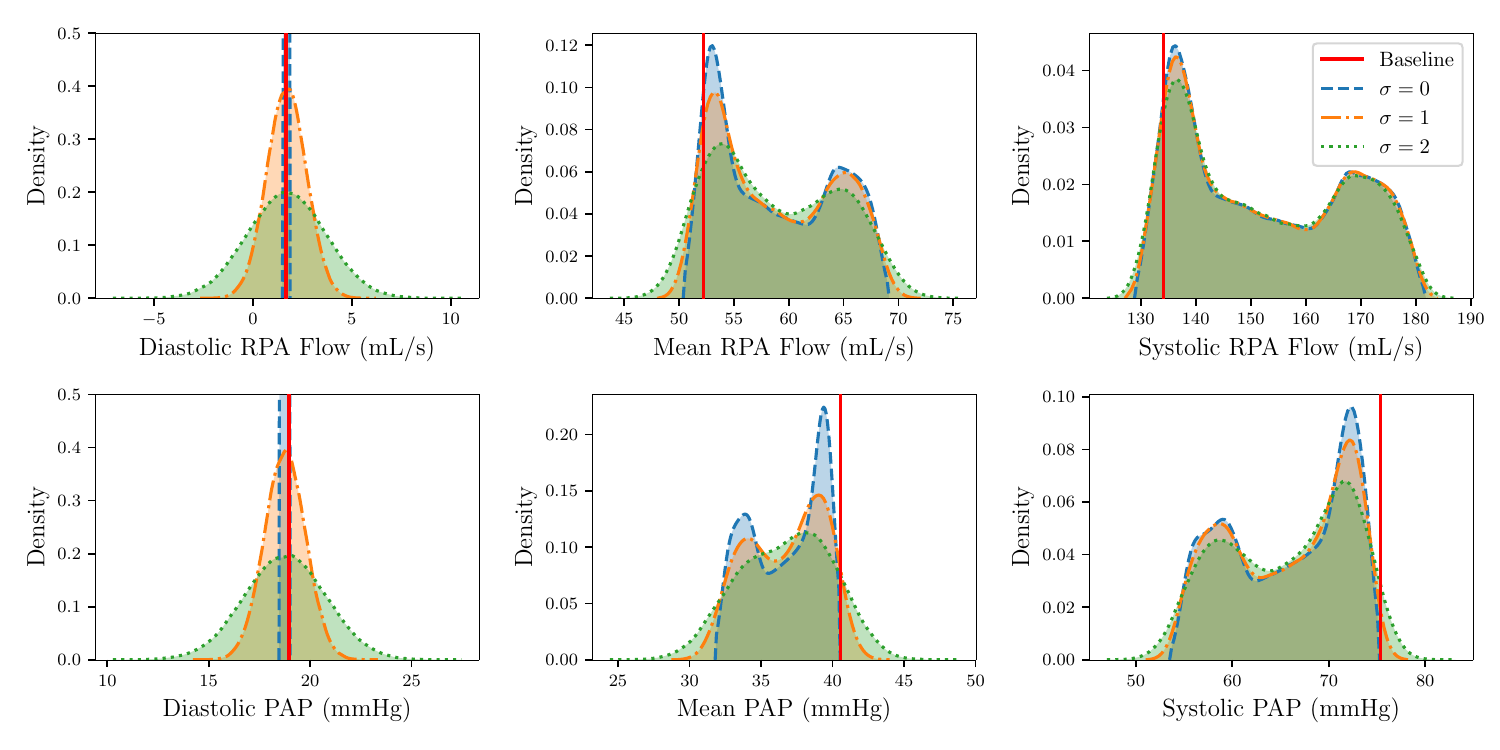}
\caption{Marginal post-operative distributions of RPA flow (top row) and pulmonary artery pressure (bottom row) for different levels of catheter measurement uncertainty. A $\sigma$ of $1$ indicates $\sigma_{p} = 1$  mmHg and $\sigma_{f} = 1$ mL/s. }\label{fig:sampling_results}
\end{figure}

\subsection{Sampling}

\noindent To simulate virtual treatment, we sample various configurations $\bm{c}$ according to the hierarchical multinomial density described in \eqref{equ:c_density}. Specifically, we let $P_{F} = .3$, $P_{M} = .4$, $P_{S} = .3$, and $u_{F} = 0$, $l_{F} = u_{M} = .2$, $l_{M} = u_{S} = .8$, and $l_{S} = 1$. 
While these values have been selected for demonstration purposes, in practice they should be determined by clinicians based on their experience on repair success rates for pulmonary stenosis on specific patient cohorts.

In addition, the catheter measurement uncertainty $\sigma_{p}^2$ and $\sigma_{f}^2$ varies across measurement technique and instrument quality.
In Figure~\ref{fig:sampling_results}, we observe the effects of setting $\sigma_{p}$ and $\sigma_{f}$ equal to $\{0, 1, 2\}$ mmHg and mL/sec, respectively. 
As the measurement standard deviation grows larger, post-treatment pressure and flow densities progressively spread out and eventually become Gaussian at high $\sigma$, limiting the effectiveness of the proposed pipeline. In our demonstrations, we included no measurement uncertainty for clarity.

\subsection{Pre-operative analysis}\label{sec:res_pre_surgery}

\begin{figure}[ht!]
\begin{subfigure}[b]{\linewidth}
    \includegraphics[width=\textwidth]{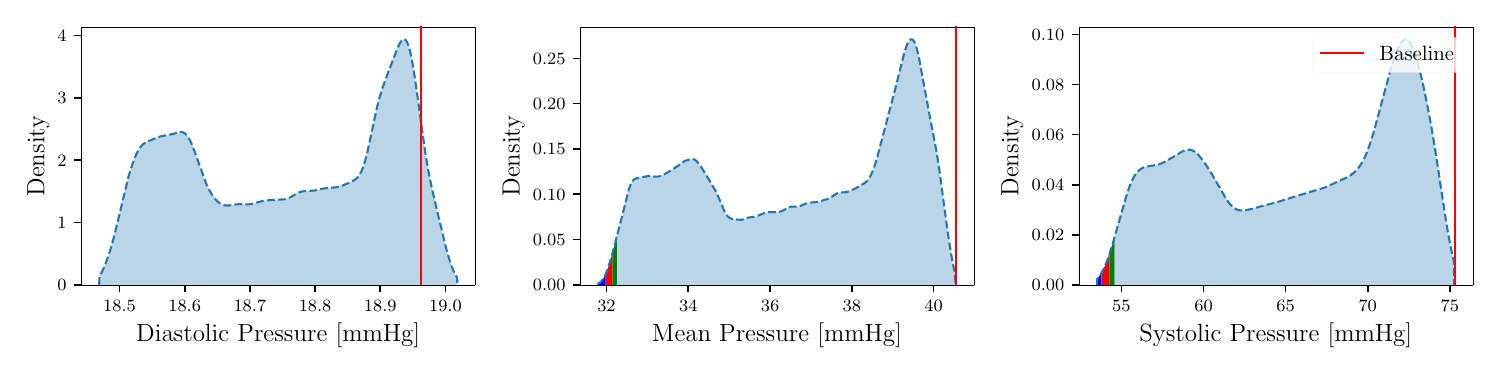}
    \caption{Histograms of marginal main pulmonary artery pressures following treatment. The vertical line in red indicates the baseline pre-operative model main pulmonary pressure.}
    \label{fig:joint_dist_hist}
\end{subfigure}
\centering
\begin{subfigure}[b]{\linewidth}
\centering
\begin{subfigure}[b]{.4\linewidth}
    \includegraphics[width=\textwidth]{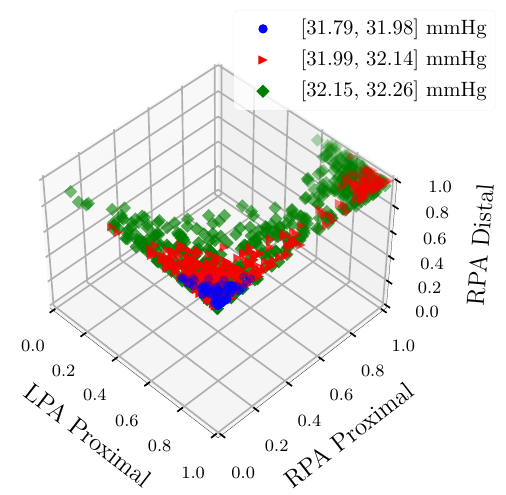}
\end{subfigure}
\begin{subfigure}[b]{.4\linewidth}
    \includegraphics[width=\textwidth]{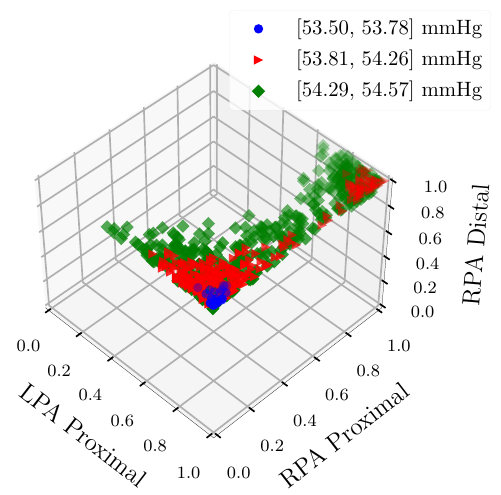}
\end{subfigure}
\caption{Repair configurations $c$ achieving the maximum possible reduction in the main pulmonary pressure. Points correspond to the lowest $0.1\%$ (blue), $0.5\%$ (red), and $1\%$ (green) of all possible pressure realizations.}
\label{fig:joint_dist_dens}
\end{subfigure}
\caption{Pre-treatment analysis of main pulmonary pressures for model AS1\_SU0308.}
\label{fig:joint_dist}
\end{figure}
\noindent Pre-operatively, the proposed framework can be used to answer questions related to the range of patient-specific flows and pressures that are achievable in the short-term following treatment. 
To answer this question, we generate roughly $N$=100,000 samples from $\bm{c}$ according to the density in~\eqref{equ:c_density}, run them through the neural twin, and extract pulmonary artery pressure (PAP) outcomes, as shown in Figure~\ref{fig:joint_dist_hist}. 
On the same GPU used to train the neural twin, sampling took only 9 seconds.
For any pressure targets, we can then answer the above question through a frequentist estimate of the probability $\rho(p_l < p < p_u)$.
In practice, however, a clinician may wish to identify hemodynamic outcomes at a number of locations in the vasculature or, for example, improve flow balance between the RPA and LPA. 

Analyzing the $n$-dimensional density of configurations $\bm{c}$ corresponding to the solution space of the previous problem can provide an answer to the second question.
Figure~\ref{fig:joint_dist_dens} plots the spatial density of $\bm{c}$ for pressures in the lowest $0.1\%$, $0.5\%$, and $1\%$ of pressures, roughly aggregated as the target region $31.79 < p <  32.26$ mmHg.
This figure suggests how the complete repair of the RPA distal stenosis is crucial, followed by additional marginal pressure reduction by fixing the LPA proximal region. 
Interestingly, we note that although the effects of fixing the RPA proximal lesion are minimal, doing so is actually detrimental to our goal.

While the analysis of repair importance provides a general overview of repair target priority, it fails to account for how probable achieving particular configurations is. Thus, it is insufficient as an answer to our final question - optimal repair order.

Firstly, the answer to such a question depends heavily on the methodology for quantifying repair success, which in turn relies upon the clinically identified needs of an individual patient. Certain repair orders may introduce possibilities for dangerous situations such as excessive pulmonary perfusion or, as we observe in the next section, remove the need to repair certain vessels. Additionally, with the combinatoric nature of computing conditional and marginals for an $n-dimensional$ problem, the computational feasibility of simulation algorithms must also be considered. Instead of formulating the simplified pre-operative problem, which is rather uninteresting given only $n=3$, we instead turn our attention to how the answer to question 3 might change during treatment (either transcatheter pulmonary artery rehabilitation or surgery).

\subsection{Peri-operative analysis}\label{sec:res_intra_surgery}
\begin{figure}[ht!]
\centering
\begin{subfigure}[b]{.45\linewidth}
    \includegraphics[width=\textwidth]{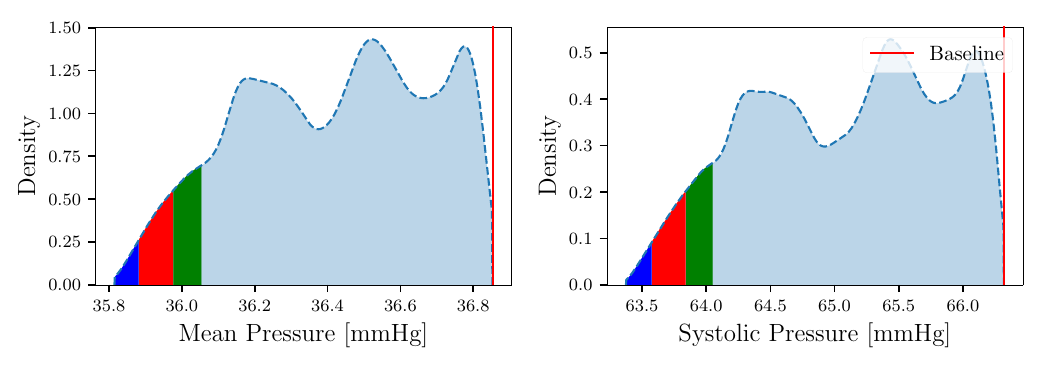}
    \caption{Conditional probability of the mean and systolic pulmonary pressure given $c_{2} = 0.6$.}
    \label{fig:cond_hist_1}
\end{subfigure}
\begin{subfigure}[b]{.45\linewidth}
    \includegraphics[width=\textwidth]{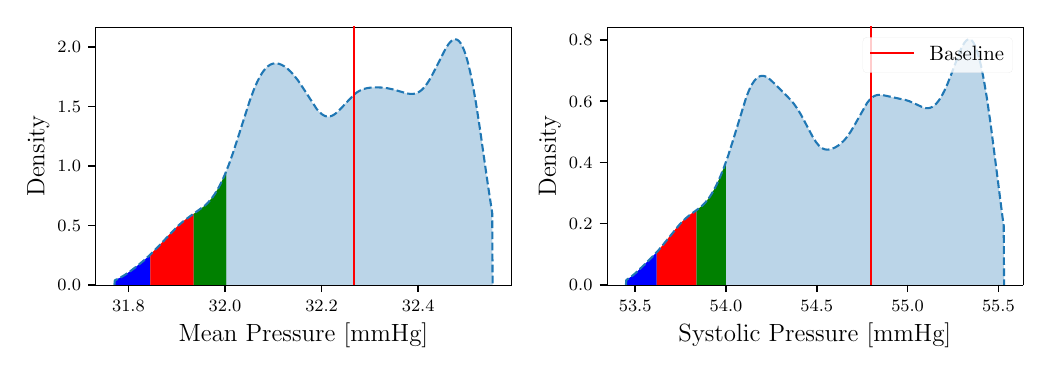}
    \caption{Conditional probability of the mean and systolic pulmonary pressure given $c_{2} = 1$.}
    \label{fig:cond_hist_2}
\end{subfigure}
\centering
\begin{subfigure}[b]{.45\linewidth}
\centering
    \begin{subfigure}[b]{.47\linewidth}
    \includegraphics[width=\textwidth]{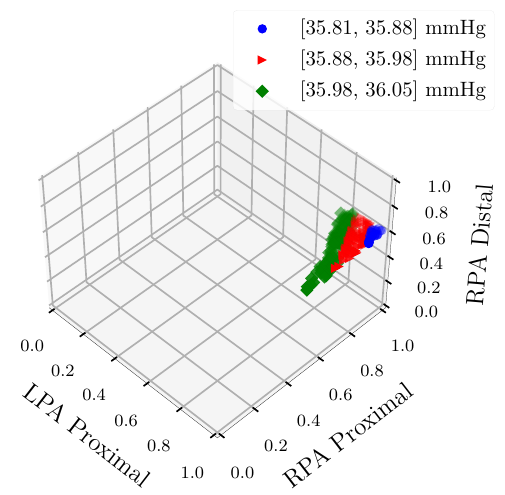}
   
    \end{subfigure}
    \begin{subfigure}[b]{.47\linewidth}
    \includegraphics[width=\textwidth]{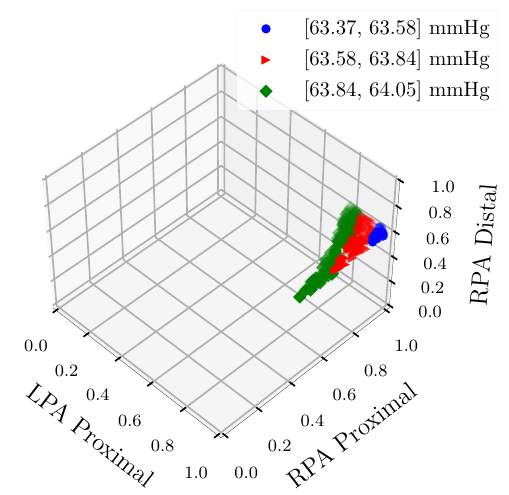}
    \end{subfigure}
    \caption{Degrees of repair leading to the maximum reduction in the mean pulmonary pressure, conditioned to $c_3 = 0.6$.}
\label{fig:cond_density_1}
\end{subfigure}
\begin{subfigure}[b]{.45\linewidth}
\centering
    \begin{subfigure}[b]{.47\linewidth}
    \includegraphics[width=\textwidth]{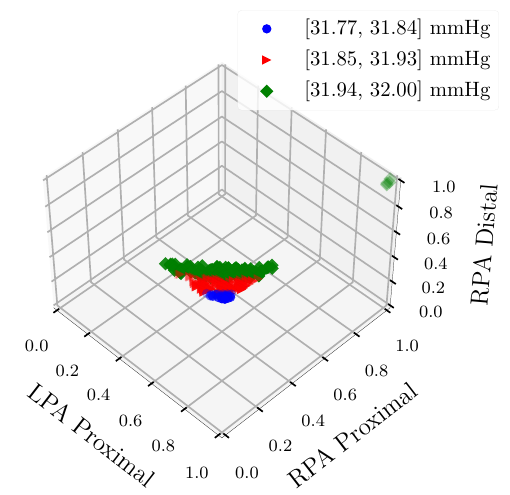}
    \end{subfigure}
    \begin{subfigure}[b]{.47\linewidth}
    \includegraphics[width=\textwidth]{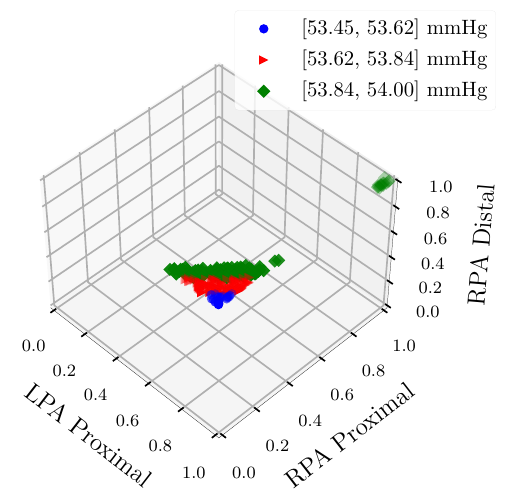}
    \end{subfigure}
    \caption{Degrees of repair leading to the maximum reduction in the mean pulmonary pressure, conditioned to $c_3 = 1$.}
\label{fig:cond_density_2}
\end{subfigure}
\label{fig:lower_bins_density}
\caption{Peri-operative analysis of peripheral pulmonary stenosis repair. Baselines in (a) and (b) are determined by the realized repair, or $\bm{c} =  <0,0,0.6>$ and  $\bm{c} = <0,0,1>$ respectively. Density plots in (c) and (d) to the lowest $1\%$ (blue), $5\%$ (red), and $10\%$ (green) of pressures reachable given their respective conditions.  }
\end{figure}
\noindent As a clinician performs balloon angioplasty or implants a stent at the $i$-th designated repair location, a realization becomes available for the random variable $c_i$. This changes the nature of the sought result from a joint to a conditional probability, given the amount of actual peri-operative repair.
In Figures~\ref{fig:cond_hist_1} and~\ref{fig:cond_hist_2}, we sample another 100,000 configurations of $\bm{c}$, assuming the RPA distal repair is fixed to $0.6$ ($c_3 = 0.6$) and $1.0$ ($c_3 = 1.0$), respectively. 
When $c_3 = 0.6$, we note that achieving our previously desired pressure targets is actually no longer possible, confirming the utility of online predictions.
Additionally, Figure~\ref{fig:cond_density_1} and~\ref{fig:cond_density_2} show the configurations $c$ corresponding to minimal main pulmonary pressures for RPA distal repairs fixed at either 0.6 or 1.0.
At $c_3 = 0.6$, we can achieve further minimization of pressures by repairing both the LPA and RPA proximal regions, whereas at $c_3 = 1.0$, repairing the RPA proximal locations seems unnecessary.  

\section{Discussion}\label{sec:discussion}

\noindent This paper introduces a prototype system for predicting the hemodynamic consequences of stenosis repair in peripheral pulmonary artery disease. 
The problem is expressed in probabilistic terms, and the joint probability of pressure, flow, and the degree of repair is formulated under the assumption of \emph{independent repairs} and Gaussian measurement noise. 
We proposed a novel, non-invasive solution to achieving accurate hemodynamic outputs within zero-dimensional models, performing iterative linear corrections to optimize resistances and account for pressure losses within junctions. The method, although particularly useful in branching pulmonary models, serves as a general technique for mitigating 0D-3D discrepancies in any vasculature.
We also proposed an original solution to the parameterization of arbitrarily shaped stenotic lesions through a zero-dimensional representation of desired hemodynamic states resulting from iterative linear corrections. 
We then train a fast, fully-connected neural emulator to determine the hemodynamic response following repair and investigate its use in pre-treatment planning and for real-time decision making during the intervention.

Although the proposed approach is effective and applicable to general anatomies and an arbitrary number of lesions, there are a few limitations and modifications worth addressing. 
First, in Section~\ref{sec:bc_tuning}, we presented one scheme for boundary condition tuning, inherently limited by the available physiological information. However, our scheme should be interchangeable with any other tuning framework, attending to different available physiological targets. Moreover, boundary conditions are tuned for the short-term treatment follow-up and disregard growth, remodeling, and adaptation~\cite{yang2016adaptive}. 

In Section~\ref{sec:linear_correction} we presented our approach only correcting linear resistors and targeting the mean pressures, leading to a possible underestimation of main pulmonary artery pressures at systole and reduced accuracy in matching the diastolic pressures. However, any pressure or flow at any location in the model and time during the heart cycle could serve as a target for linear correction. Additionally, corrections could also be applied to capacitors, inductors, and non-linear resistors. 

In Section~\ref{sec:parameterization}, we combine repaired hemodynamic states based on independence and assuming linear superposition, which may not account for non-linear interaction between stenosis repairs, particularly in extended stenotic regions close to each other. We also acknowledge that the current method of determining the locality of stenosis is particularly involved, requiring manual inspection of centerline deviation. More formulaic, automated methods could be utilized to determine the impact region of a stenosis.

Finally, neural emulators were constructed exclusively using 0D models and are fundamentally limited in true hemodynamic accuracy. Implementing multi-fidelity model fusion should enable neural emulators to maintain high accuracy while continuing to provide sufficient data to capture changes corresponding to virtual repairs.

In this study, we explore the usage of cardiovascular digital twins as a treatment planning aid, seeking to minimize risks associated with intervention and improve clinician insights into the patient-specific procedure. Our pipeline serves as the foundation and direction for new studies to improve the utility of treatment planning tools. Future research can rigorously address the limitations we acknowledged, or discover new methods to answer relevant questions using our neural surrogate. 

\section*{Acknowledgements}\label{sec:acks}

\noindent JDL acknowledges the support from two REU supplements for NSF CAREER award \#1942662 (PI DES) during Summer 2021 at the University of Notre Dame, and a supplement for NSF CDS\&E \#2105345 (PI ALM) during Summer 2022 at Stanford University.
DES was supported by a NSF CAREER award \#1942662 and a NSF CDS\&E award \#2104831. DES, JK, JAF and ALM acknowledge support from NIH grant \#1R01HL167516 \emph{Uncertainty aware virtual treatment planning for peripheral pulmonary artery stenosis} (PI ALM). MRP acknowledges the support of NIH R01 Grant \#K99HL161313 and the Stanford Maternal and Child Health Research Institute.
The authors would like to thank Stanford University, the Stanford Research Computing Center, and the Center for Research Computing at the University of Notre Dame for providing computational resources and support that were essential to generate model results for this study.

\section*{Ethical statement}

\noindent None

\bibliographystyle{abbrv}
\bibliography{ref}

\begin{thebibliography}{10}

\bibitem{antiga2008image}
L.~Antiga, M.~Piccinelli, L.~Botti, B.~Ene-Iordache, A.~Remuzzi, and D.~A.
  Steinman.
\newblock An image-based modeling framework for patient-specific computational
  hemodynamics.
\newblock {\em Medical \& biological engineering \& computing}, 46:1097--1112,
  2008.

\bibitem{Bjornsson_2019}
B.~Björnsson, C.~Borrebaeck, N.~Elander, T.~Gasslander, D.~R. Gawel,
  M.~Gustafsson, R.~Jörnsten, E.~J. Lee, X.~Li, S.~Lilja, and et~al.
\newblock Digital twins to personalize medicine.
\newblock {\em Genome Medicine}, 12(1), 2019.

\bibitem{Coelho_2020}
G.~Coelho, N.~N. Rabelo, E.~Vieira, K.~Mendes, G.~Zagatto, R.~Santos~de
  Oliveira, C.~E. Raposo-Amaral, M.~Yoshida, M.~R. de~Souza, C.~F. Fagundes,
  and et~al.
\newblock Augmented reality and physical hybrid model simulation for
  preoperative planning of metopic craniosynostosis surgery.
\newblock {\em Neurosurgical Focus}, 48(3), Mar 2020.

\bibitem{Collins_2019}
R.~T. Collins, R.~D. Mainwaring, K.~L. MacMillen, and F.~L. Hanley.
\newblock Outcomes of pulmonary artery reconstruction in {W}illiams syndrome.
\newblock {\em The Annals of Thoracic Surgery}, 108(1):146–153, Mar 2019.

\bibitem{Cunningham_2013}
J.~W. Cunningham, D.~B. McElhinney, K.~Gauvreau, L.~Bergersen, R.~V. Lacro,
  A.~C. Marshall, L.~Smoot, and J.~E. Lock.
\newblock Outcomes after primary transcatheter therapy in infants and young
  children with severe bilateral peripheral pulmonary artery stenosis.
\newblock {\em Circulation: Cardiovascular Interventions}, 6(4):460–467, Aug
  2013.

\bibitem{figueroa2006coupled}
C.~A. Figueroa, I.~E. Vignon-Clementel, K.~E. Jansen, T.~J. Hughes, and C.~A.
  Taylor.
\newblock A coupled momentum method for modeling blood flow in
  three-dimensional deformable arteries.
\newblock {\em Computer methods in applied mechanics and engineering},
  195(41-43):5685--5706, 2006.

\bibitem{Geggel_Gauvreau_Lock_2001}
R.~L. Geggel, K.~Gauvreau, and J.~E. Lock.
\newblock Balloon dilation angioplasty of peripheral pulmonary stenosis
  associated with {W}illiams syndrome.
\newblock {\em Circulation}, 103(17):2165–2170, May 2001.

\bibitem{gelman2006data}
A.~Gelman and J.~Hill.
\newblock {\em Data analysis using regression and multilevel/hierarchical
  models}.
\newblock Cambridge university press, 2006.

\bibitem{Hallbergson_2013}
A.~Hallbergson, J.~E. Lock, and A.~C. Marshall.
\newblock Frequency and risk of in-stent stenosis following pulmonary artery
  stenting.
\newblock {\em The American Journal of Cardiology}, 113(3):541–545, Nov 2013.

\bibitem{Inglessis_Landzberg_2007}
I.~Inglessis and M.~J. Landzberg.
\newblock Interventional catheterization in adult congenital heart disease.
\newblock {\em Circulation}, 115(12):1622–1633, 2007.

\bibitem{virtual_intervention}
I.~S. Lan, W.~Yang, J.~A. Feinstein, J.~Kreutzer, R.~T. Collins, M.~Ma, G.~T.
  Adamson, and A.~L. Marsden.
\newblock Virtual transcatheter interventions for peripheral pulmonary artery
  stenosis in {W}illiams and {A}lagille syndromes.
\newblock {\em Journal of the American Heart Association}, 11(6), 2022.

\bibitem{liu2020fluid}
J.~Liu, W.~Yang, I.~S. Lan, and A.~L. Marsden.
\newblock Fluid-structure interaction modeling of blood flow in the pulmonary
  arteries using the unified continuum and variational multiscale formulation.
\newblock {\em Mechanics research communications}, 107:103556, 2020.

\bibitem{Luong_2020}
R.~Luong, J.~A. Feinstein, M.~Ma, N.~H. Ebel, L.~Wise-Faberowski, Y.~Zhang,
  L.~F. Peng, V.~V. Yarlagadda, J.~Shek, F.~L. Hanley, and et~al.
\newblock Outcomes in patients with {A}lagille syndrome and complex pulmonary
  artery disease.
\newblock {\em The Journal of Pediatrics}, 229, Sep 2020.

\bibitem{meng2020composite}
X.~Meng and G.~E. Karniadakis.
\newblock A composite neural network that learns from multi-fidelity data:
  Application to function approximation and inverse {PDE} problems.
\newblock {\em Journal of Computational Physics}, 401:109020, 2020.

\bibitem{milivsic2004analysis}
V.~Mili{\v{s}}i{\'c} and A.~Quarteroni.
\newblock Analysis of lumped parameter models for blood flow simulations and
  their relation with {1D} models.
\newblock {\em ESAIM: Mathematical modelling and numerical analysis},
  38(4):613--632, 2004.

\bibitem{murray1926physiological}
C.~D. Murray.
\newblock The physiological principle of minimum work applied to the angle of
  branching of arteries.
\newblock {\em The Journal of general physiology}, 9(6):835, 1926.

\bibitem{nelder1965simplex}
J.~A. Nelder and R.~Mead.
\newblock A simplex method for function minimization.
\newblock {\em The computer journal}, 7(4):308--313, 1965.

\bibitem{automated_0d}
M.~R. Pfaller, J.~Pham, A.~Verma, L.~Pegolotti, N.~M. Wilson, D.~W. Parker,
  W.~Yang, and A.~L. Marsden.
\newblock Automated generation of {0D} and {1D} reduced‐order models of
  patient‐specific blood flow.
\newblock {\em International Journal for Numerical Methods in Biomedical
  Engineering}, 2022.

\bibitem{Shu_2023}
H.~Shu, R.~Liang, Z.~Li, A.~Goodridge, X.~Zhang, H.~Ding, N.~Nagururu, M.~Sahu,
  F.~X. Creighton, R.~H. Taylor, and et~al.
\newblock Twin-s: A digital twin for skull base surgery.
\newblock {\em International Journal of Computer Assisted Radiology and
  Surgery}, 18(6):1077–1084, 2023.

\bibitem{sobol1967distribution}
I.~M. Sobol'.
\newblock On the distribution of points in a cube and the approximate
  evaluation of integrals.
\newblock {\em Zhurnal Vychislitel'noi Matematiki i Matematicheskoi Fiziki},
  7(4):784--802, 1967.

\bibitem{Taylor_1994}
R.~Taylor, B.~Mittelstadt, H.~Paul, W.~Hanson, P.~Kazanzides, J.~Zuhars,
  B.~Williamson, B.~Musits, E.~Glassman, and W.~Bargar.
\newblock An image-directed robotic system for precise orthopaedic surgery.
\newblock {\em IEEE Transactions on Robotics and Automation}, 10(3):261–275,
  Jun 1994.

\bibitem{TRIVEDI_BENSON_2003}
K.~R. TRIVEDI and L.~N. BENSON.
\newblock Interventional strategies in the management of peripheral pulmonary
  artery stenosis.
\newblock {\em Journal of Interventional Cardiology}, 16(2):171–188, Apr
  2003.

\bibitem{updegrove2017simvascular}
A.~Updegrove, N.~M. Wilson, J.~Merkow, H.~Lan, A.~L. Marsden, and S.~C.
  Shadden.
\newblock Simvascular: an open source pipeline for cardiovascular simulation.
\newblock {\em Annals of biomedical engineering}, 45:525--541, 2017.

\bibitem{whiting2001stabilized}
C.~H. Whiting and K.~E. Jansen.
\newblock A stabilized finite element method for the incompressible
  {N}avier--{S}tokes equations using a hierarchical basis.
\newblock {\em International Journal for Numerical Methods in Fluids},
  35(1):93--116, 2001.

\bibitem{yang2016adaptive}
W.~Yang, J.~A. Feinstein, and I.~E. Vignon-Clementel.
\newblock Adaptive outflow boundary conditions improve post-operative
  predictions after repair of peripheral pulmonary artery stenosis.
\newblock {\em Biomechanics and modeling in mechanobiology}, 15:1345--1353,
  2016.

\end{thebibliography}

\appendix

\section{Addendum on Boundary Condition Tuning}\label{sec:app_nonlin_res}

\noindent In this appendix, we detail additional steps for our boundary tuning procedure, covering flow waveform adjustment, capillary wedge pressure modification, and the process for determining non-linear resistances.

As mentioned in the paper, blood flow for model AS\_SU0308 was not originally provided as a part of clinical data. To overcome this problem, in~\cite{virtual_intervention}, Lan \textit{et. al.} constructed a closed-loop right-heart model, generating a realistic flow waveform by optimizing for available physiological targets. Upon retrieving the results, we found that while the flow waveform adequately matched systolic, mean, and diastolic pressures, it underrepresented the expected cardiac output (5.32 L/min measured versus 5.838 L/min expected). To correct for this difference without changing diastolic or systolic pressures, we added two disjoint sinusoidal terms as follows.

\begin{figure}[ht!]\newcommand\tw{0.065}
\centering
\begin{subfigure}[t]{0.5\linewidth}
    \includegraphics[width=\textwidth]{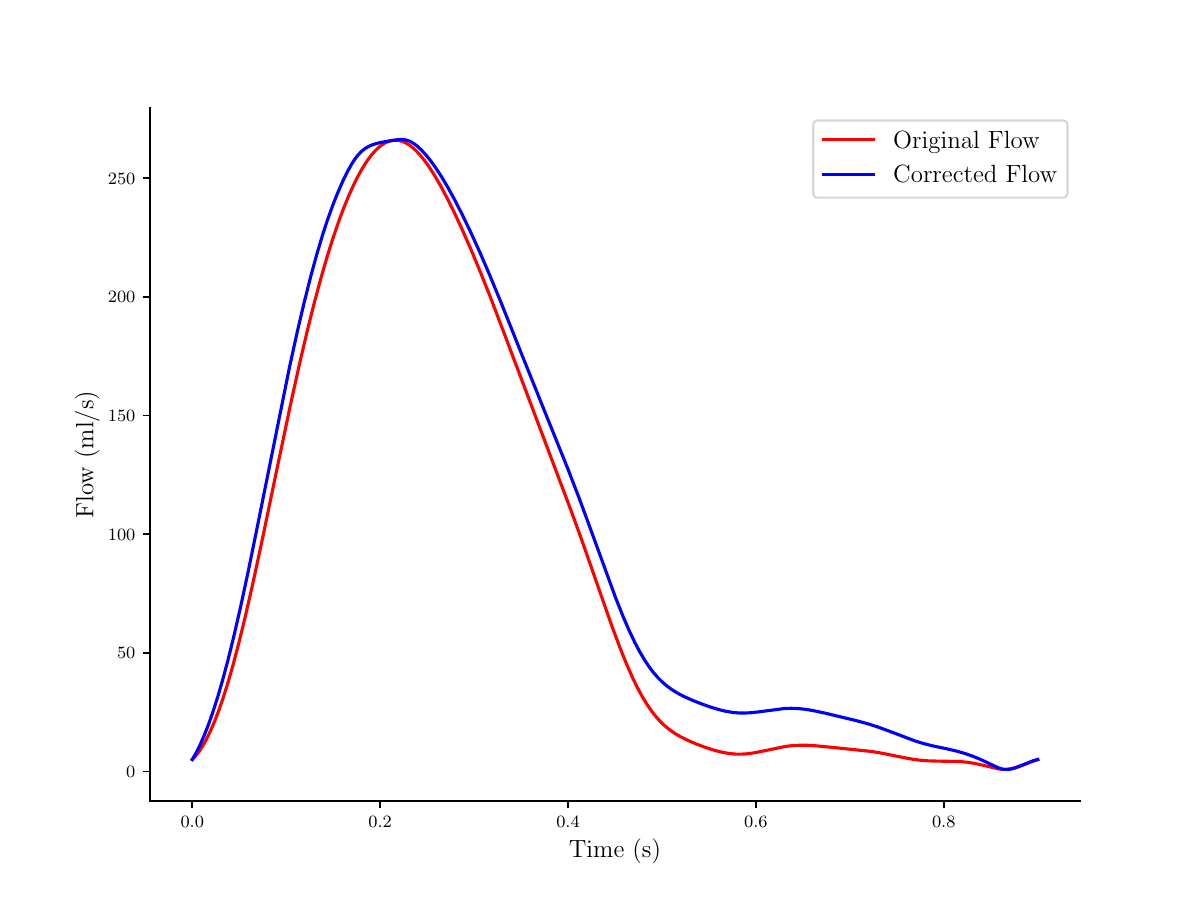}
\end{subfigure}
\caption{Corrected versus original inflow waveform (VMR ID: AS1\_SU0308).}
\label{fig:corrected_inflow}
\end{figure}

First, we determined the difference in cardiac output, $\Delta CO$, and the time points at start, systole, and diastole, denoted $t_0$, $t_{\mathrm{sys}}$ and $t_{\mathrm{dia}}$, respectively. We then split $\Delta CO$ into two terms scaled by time, $\Delta CO_1$ and $\Delta CO_2$, computed as follows
\begin{equation}
    \Delta CO_1 = \Delta CO \cdot \frac{t_{\mathrm{sys}} - t_0}{t_{\mathrm{dia}} - t_0}, \quad
    \Delta CO_2 = \Delta CO \cdot \frac{t_{\mathrm{dia}} - t_{\mathrm{sys}}}{t_{\mathrm{dia}} - t_0}.
\end{equation}
We then computed two sinusoids $f(t)$ and $g(t)$ on $[t_0, t_{\mathrm{sys}}]$ and $[t_{\mathrm{sys}}, t_{\mathrm{dia}}]$ respectively, such that $f(t_0) = f(t_{\mathrm{sys}}) = g(t_{\mathrm{sys}}) = g(t_{\mathrm{dia}}) = 0$. The equations for the sinusoids are
\begin{equation}
    f(x) = \Delta CO_1 \cdot \frac{\pi}{2(t_{\mathrm{sys}} - t_0)} \cdot \sin\left(\frac{\pi\,t}{t_{\mathrm{sys}} - t_0}\right), \quad g(x) = \Delta CO_2 \cdot \frac{\pi}{2(t_{\mathrm{dia}} - t_{\mathrm{sys}})} \cdot \sin\left(\frac{\pi\,t}{t_{\mathrm{dia}} - t_{\mathrm{sys}}}\right)
\end{equation}
After adding each sinusoidal term to their respective regions, the final measured cardiac output was 5.89 L/min. Figure~\ref{fig:corrected_inflow} compares the uncorrected and corrected inflow waveforms. 

The capillary wedge pressure for model AS1\_SU0308 was measured to have an average of $20$ mmHg, whereas the diastolic pressure was measured at $18$ mmHg. In zero-dimensional models, the governing equations disallow backflow. Therefore, if we set our constant atrial pressure to be $20$ mmHg, it would be impossible to tune parameters within our tuning model such that diastolic pressure is lower than $20$ mmHg. We note that to correct this issue, throughout the models in our paper, we limit the atrial pressure to be $18$ mmHg.

The non linear resistors for boundary condition tuning were determined as follows. We first determined resistance boundary conditions for a steady state simulation. Specifically, we determined a total boundary condition resistance as $R_{\text{tot}} = $(mPAP-LAP)/CO$ = ((42-0)\cdot 1333.22)/96.3$ = 581 Barye$\cdot$s/ml. This total resistance was distributed among outlets using Murray's law~\cite{murray1926physiological}. Based on the results of the simulation we estimated the LPA/RPA and total pulmonary resistance in the three dimensional models as reported in Table~\ref{table:steady_co}. Branch and total resistance was estimated following two different approaches to assess variability.
In the first approach (mean pressure drop) we computed the pressure drops between the inlet and each pulmonary outlet and divided their mean value by the cardiac output. In the second approach, an equivalent resistance was computed by assuming all outlet resistances to be in parallel. 
To check how this resistance was affected by the inlet flow, we quantified diastolic, mean and systolic flow as 0.120, 5.8 and 19.8 L/min, respectively, and conducted steady state simulations, with results reported in Table~\ref{table:steady_multiple}.
Finally, model resistance from various steady flows were computed in Table~\ref{table:steady_res} assuming a parallel arragement and with an arthmetic or weighted average. The plots in Figure~\ref{fig:Q_vs_R} were generated using this table.

\begin{table}[!ht]
\centering
\caption{Estimating three-dimensional model resistance from steady state simulation with CO inflow.}\label{table:steady_co}
\begin{tabular}{l c c} 
\toprule
{\bf Qty} & {\bf Value} & {\bf Units}\\
\midrule
Mean pressure drop across LPA & 5.4 & mmHg\\
Mean pressure drop across RPA & 8.5 & mmHg\\
\midrule
Flow LPA & 2.9 & L/min\\
Flow RPA & 2.5 & L/min\\
RPA Flow Split & 46.2 & \%\\
Total Pulmonary Flow & 5.8 & L/min\\
\midrule
Total LPA Resistance (Mean pressure drop) & 150.3 & Barye$\cdot$s/ml\\
Total LPA Resistance (Parallel) & 145.0 & Barye$\cdot$s/ml\\
\midrule
Total RPA Resistance (Mean pressure drop) & 276.4 & Barye$\cdot$s/ml\\
Total RPA Resistance (Parallel) & 192.0 & Barye$\cdot$s/ml\\
\midrule
Total Pulmonary Resistance (Mean pressure drop) & 97.4 & Barye$\cdot$s/ml\\
Total Pulmonary Resistance (Parallel) & 82.6 & Barye$\cdot$s/ml\\
\bottomrule
\end{tabular}
\end{table}

\begin{table}[!ht]
\centering
\caption{Model results from steady state simulations with distolic, mean and systolic flow.}
\begin{tabular}{l c c c c} 
\toprule
{\bf Qty} & {\bf DIA} & {\bf MEAN} & {\bf SYS} & {\bf Units}\\
\midrule
Mean pressure drop across LPA & 0.017 & 6.3 & 73.4 & mmHg\\
Mean pressure drop across RPA & 0.023 & 9.9 & 93.2 & mmHg\\
\midrule
Flow LPA & 0.062 & 3.1 & 10.9 & L/min\\
Flow RPA & 0.058 & 2.7 & 8.9 & L/min\\
RPA Flow Split & 47.932 & 46.1 & 44.9 & [\%]\\
Total Pulmonary Flow & 0.120 & 5.8 & 19.8 & L/min\\
\bottomrule
\end{tabular}\label{table:steady_multiple}
\end{table}

\begin{table}[!ht]
\centering
\caption{Estimation of model resistance from distolic, mean and systolic steady flow.}
\begin{tabular}{l c c c c c} 
\toprule
{\bf DIASTOLIC} & {\bf Par} & {\bf Avg} & {\bf Area} & {\bf Inv. Area} & {\bf Units}\\
\midrule
Total LPA Resistance & 19.5 & 21.7 & 20.4 & 22.7 & Barye$\cdot$s/ml\\
Total RPA Resistance & 29.7 & 32.5 & 32.1 & 33.1 & Barye$\cdot$s/ml\\
Total Pulmonary Resistance & 11.8 & 13.0 & 12.5 & 13.5 & Barye$\cdot$s/ml\\
\midrule
{\bf MEAN} & {\bf Par} & {\bf Avg} & {\bf Area} & {\bf Inv. Area} & {\bf Units}\\
\midrule
Total LPA Resistance & 156.7 & 162.0 & 164.3 & 160.4 & Barye$\cdot$s/ml\\
Total RPA Resistance & 204.8 & 297.4 & 297.2 & 304.2 & Barye$\cdot$s/ml\\
Total Pulmonary Resistance & 88.8 & 104.9 & 105.8 & 105.0 & Barye$\cdot$s/ml\\
\midrule
{\bf SYSTOLIC} & {\bf Par} & {\bf Avg} & {\bf Area} & {\bf Inv. Area} & {\bf Units}\\
\midrule
Total LPA Resistance & 525.2 & 538.2 & 548.1 & 531.3 & Barye$\cdot$s/ml\\
Total RPA Resistance & 494.6 & 838.8 & 841.8 & 853.5 & Barye$\cdot$s/ml\\
Total Pulmonary Resistance & 254.7 & 327.8 & 332.0 & 327.5 & Barye$\cdot$s/ml\\
\bottomrule
\end{tabular}\label{table:steady_res}
\end{table}

\end{document}